\documentclass[12pt]{article}
\usepackage{geometry}
\usepackage[onehalfspacing]{setspace}
\usepackage{amssymb}
\usepackage{amsfonts}
\usepackage{graphicx}
\usepackage{amsmath}
\usepackage{epstopdf}
\usepackage{natbib}
\usepackage{bibunits}
\usepackage{morefloats}
\usepackage{float}
\usepackage{comment}
\usepackage{pdflscape}
\usepackage{adjustbox}
\usepackage{subcaption}
\usepackage{longtable}
\usepackage{threeparttablex}
\usepackage{hhline,multirow}
\usepackage[dvipsnames,table,xcdraw]{xcolor}
\usepackage{txfonts}
\usepackage{multicol}
\usepackage{booktabs}
\usepackage{multicol}
\usepackage{booktabs}
\usepackage{threeparttable}
\usepackage{tabularx}
\usepackage{dcolumn}
\usepackage[final]{pdfpages}
\usepackage{dcolumn}
\usepackage{threeparttable}
\usepackage{tabularx}
\usepackage{mathpazo}
\usepackage[scaled]{helvet}
\usepackage{courier}
\usepackage[T1]{fontenc}
\usepackage[pdftex,bookmarks,colorlinks]{hyperref}
\usepackage[compact]{titlesec}
\usepackage{enumitem}
\usepackage[bottom]{footmisc}

\newcolumntype{d}[1]{D..{#1}} 
\normalfont
\hypersetup{
  colorlinks,
  citecolor=blue,
  linkcolor=red,
  urlcolor=Green}
\titlespacing{\section}{0pt}{0pt}{8pt}
\titlespacing{\subsection}{0pt}{2pt}{5pt}
\titlespacing{\subsubsection}{0pt}{3pt}{5pt}
\setlist[enumerate]{nosep,topsep={2pt},partopsep={2pt}}

\renewenvironment{abstract}
 {\small
  \begin{center}
  \bfseries \abstractname\vspace{-.5em}\vspace{0pt}
  \end{center}
  \list{}{
    \setlength{\leftmargin}{3.1cm}    \setlength{\rightmargin}{\leftmargin}  }  \item\relax}
 {\endlist}
\begin{document}

\title{{\LARGE \textbf{Can Machine Learning Catch the COVID-19 Recession?%
\thanks{%
We thank the Editor Ana Galvao, an anonymous referee, and Hugo Couture who
provided excellent research assistance. The third author acknowledges
financial support from the Chaire en macro\'{e}conomie et pr\'{e}visions ESG
UQAM.}}}}
\author{Philippe Goulet Coulombe \thanks{%
University of Pennsylvania} \hspace{5mm} Massimiliano Marcellino \thanks{%
Bocconi University, IGIER, Baffi-Carefin, BIDSA and CEPR} \hspace{5mm}
Dalibor Stevanovi\'{c} \thanks{%
Universit\'{e} du Qu\'{e}bec \`{a} Montr\'{e}al and CIRANO}}
\date{ \today}
\maketitle

\begin{abstract}
\noindent Based on evidence gathered from a newly built large macroeconomic
data set for the UK, labeled UK-MD and comparable to similar datasets for
the US and Canada, it seems the most promising avenue for forecasting during
the pandemic is to allow for general forms of nonlinearity by using machine
learning (ML) methods. But not all nonlinear ML methods are alike. For
instance, some do not allow to extrapolate (like regular trees and forests)
and some do (when complemented with linear dynamic components). This and
other crucial aspects of ML-based forecasting in unprecedented times are
studied in an extensive pseudo-out-of-sample exercise.
\end{abstract}

\vspace{1.5cm}

\thispagestyle{empty}



\clearpage

\newgeometry{left=1.7 cm, right= 1.7 cm, top=1.7 cm, bottom=1.7 cm}




\section{Introduction}

Forecasting economic developments during crisis time is problematic since
the realizations of the variables are far away from their average values,
while econometric models are typically better at explaining and predicting
values close to the average, particularly so in the case of linear models.
The situation is even worse for the Covid-19 induced recession, when
typically well performing econometric models such as Bayesian VARs with
stochastic volatility have troubles in tracking the unprecedented fall in
real activity and labour market indicators --- see for example for the US 
\cite{Carriero-Clark-Marcellino(2020)} and \cite%
{Plagborg-Moller-Reichlin-Ricco-Hasenzagl(2020)}, or \cite{An-Loungani(2020)}
for an analysis of the past performance of the Consensus Forecasts.

As a partial solution, \cite{Foroni-Marcellino-Stevanovic(2020)} employ
simple mixed-frequency models to nowcast and forecast US and the rest of G7
GDP quarterly growth rates, using common monthly indicators, such as
industrial production, surveys, and the slope of the yield curve. They then
adjust the forecasts by a specific form of intercept correction or estimate
by the similarity approach, see \cite{Clements-Hendry(1999)} and \cite%
{Dendramis-Kapetanios-Marcellino(2020)}, showing that the former can reduce
the extent of the forecast error during the Covid-19 period. \cite%
{Schorfheide-Song(2020)} do not include COVID periods in the estimation of a
mixed-frequency VAR model because those observations substantially alter the
forecasts. An alternative approach is the specification of sophisticated
nonlinear / time-varying models. While this is not without perils when used
on short economic time series, it can yield some gains, see e.g. \cite%
{Ferrara-Marcellino-Mogliani(2015)} in the context of forecasting during the
financial crisis using Markov-Switching, threshold and other types of random
parameter models.

The goal of this paper is to go one step further in terms of model
sophistication, by considering a variety of machine learning (ML) methods
and assessing whether and to what extent they can improve the forecasts,
both in general and specifically during the Covid-19 crisis, focusing on the
UK economy that at the same time was also experiencing substantial
Brexit-related uncertainty. A related paper, but with a focus on the largest
euro area countries, is \cite{huberetal2020} who introduce Bayesian Additive
Regression Tree-VARs (BART-VARs) for Covid. They develop a nonlinear
mixed-frequency VAR framework by incorporating regression trees, and
exploiting their ability to model outliers and to disentangle the signal
from noise. Indeed, the regression trees (and even more the forests) are
able to quickly adapt to extreme observations and to disentangle the switch
in the underlying regime. Another relevant related paper is \cite{GCLSS2019}%
, which however does not include an analysis of the Covid-19 period and
focuses on the US. A third related paper, again with a focus on the US, is 
\cite{chkmp(2021)}, who consider alternative specifications of BART-VARs,
possibly with also a non-parametric specification for the time-varying
volatility, and compare their point, density and tail forecast performance
with that of large Bayesian VARs with stochastic volatility, finding often
gains, though of limited size.

In line with \cite{GCLSS2019}, we consider five nonlinear nonparametric ML
methods. Three of them have the capacity to extrapolate and two do not.
Specifically, being based on trees, boosted trees (BT) and random forests
(RF) cannot predict out-of-sample a value ($\hat{y}_{i}$) greater than the
maximal in-sample value (same goes for the minimum). This is a simple
implication of how forecasts are constructed, basically by taking \textit{%
means} over sub-samples chosen in a data-driven way. Clearly, this is an
important limitation when it comes to forecasting variables which
significantly got out of their typical range during the Pandemic (like hours
worked).\footnote{%
On the other hand, this could be seen as a foolproof preventing the model to
predict incredible values.} No such constraints bind on Macroeconomic Random
Forest (MRF), Kernel Ridge Regression (KRR), and Neural Networks (NN). By
using a linear part within the leafs, MRF can extrapolate the same way a
linear model does, while retaining the usual benefits of tree-based methods
(limited or inexistent overfitting, necessitate little to no tuning, can
cope with large data). \cite{MRF} notes that this particular feature gives
MRF an edge over RF when it comes to forecasting the (once) extreme
escalation of the unemployment rate during the Great Recession.

As mentioned, we focus on the UK and, as another contribution of the paper,
we construct a monthly large-scale macroeconomic database, labeled UK-MD,
comparable to those for the US by \cite{mccracken2016fred,mccracken2020fred}
and for Canada by \cite{fglss2018}.\footnote{%
The dataset can be found here: %
\url{http://www.stevanovic.uqam.ca/DS_UKMD.html}} Specifically, the dataset
contains 112 monthly macroeconomic and financial indicators divided into
nine categories: labour, production, retail and services, consumer and
retail price indices, producer price indices, international trade, money,
credit and interest rate, stock market and finally sentiment and leading
indicators. The starting date varies across indicators, from 1960 to 2000,
and to simplify econometric analyses we also balance the resulting panel
using an EM algorithm to impute missing values, as in \cite%
{stock2002macroeconomic} and \cite{mccracken2016fred}.

In terms of empirical results, overall ML methods can provide substantial
gains when short-term forecasting several indicators of the UK economy,
though a careful temporal and variable by variable analysis is needed. Over
the full sample, RF works particularly well for labour market variables, in
particular when augmented with a Moving Average Rotation of $X$ ($X$ being
the predictors, hence "MARX"); KRR for real activity and consumer price
inflation; LASSO or LASSO+MARX for the retail price index and its version
focusing on housing; and RF for credit variables. The gains can be sizable,
even 40-50\% with respect to the benchmark, and ML methods were particularly
useful during the Covid-19 period. Focusing on the Covid sample, it is clear
that nonlinear methods with the ability to extrapolate become extremely
competitive. And this goes both ways. For instance, certain MRFs, unlike
linear methods or simpler nonlinear ML techniques, procure important
improvements by predicting unprecedented values (for hours worked), and
avoiding immaterial cataclysms (employment and housing prices).

The rest of the paper is structured as follows. Section \ref{sec:ML}
introduces the machine learning forecasting framework. Section \ref%
{sec:models} discusses the forecasting models. Section \ref{sec:UKMD}
presents the UK-MD dataset and studies its main features. Section \ref%
{sec:empirics} discusses the set-up of the forecasting exercise. Section \ref%
{sec:results} presents and discusses the results. Section \ref%
{sec:conclusion} summarizes the key findings and concludes. Additional
details and results are presented in Appendices.

\section{Machine Learning Forecasting Framework}

\label{sec:ML}

Machine learning algorithms offer ways to approximate unknown and
potentially complicated functional forms with the objective of minimizing
the expected loss of a forecast over $h$ periods. The focus of the current
paper is to construct a feature matrix susceptible to improve the
macroeconomic forecasting performance of off-the-shelf ML algorithms. Let $%
H_{t}=\left[ H_{1t},...,H_{Kt}\right] $ for $t=1,...,T$ be the vector of
variables found in a large macroeconomic dataset, such as the FRED-MD
database of \cite{mccracken2016fred} or the UK-MD dataset described in the
next section, and let $y_{t+h}$ be our target variable. We follow \cite%
{stock2002forecasting,stock2002macroeconomic} and target average growth
rates or average differences over $h$ periods ahead 
\begin{equation}
y_{t+h}=g(f_{Z}(H_{t}))+e_{t+h}\enskip.
\end{equation}%
\noindent To illustrate this point, define $Z_{t}\equiv f_{Z}(H_{t})$ as the 
$N_{Z}$-dimensional feature vector, formed by combining several
transformations of the variables in $H_{t}$.\footnote{%
Obviously, in the context of a pseudo-out-of-sample experiment, feature
matrices must be built recursively to avoid data snooping.} The function $%
f_{Z}$ represents the data pre-processing and/or featuring engineering whose
effects on forecasting performance we seek to investigate. The training
problem for the case of no data pre-processing ($f_{Z}=I()$) is 
\begin{equation}
\underset{g\in \mathcal{G}}{\text{min}}\left\{ \sum_{t=1}^{T}\left(
y_{t+h}-g\left( H_{t}\right) \right) ^{2}+\text{pen}(g;\tau )\right\}
\end{equation}%
\noindent The function $g$, chosen as a point in the functional space $%
\mathcal{G}$, maps transformed inputs into the transformed targets. $\text{%
pen()}$ is the regularization function whose strength depends on some
vector/scalar hyperparameter(s) $\tau $.

\section{Forecasting Models}

\label{sec:models}

In this section we present the main predictive models (for a more complete
discussion, see, among other, \cite{hastie2009elements}), and some
additional, less standard, forecasting models we will consider (more details
can be found in \cite{GCLSS2019}). Table \ref{fcst_models} lists all the
models implemented in the forecasting exercise, together with their
respective input matrices $Z_{t}$.

\subsection{Main models}

\noindent \textsc{\textbf{Linear Models}.} We consider the autoregressive
model (AR), as well as the autoregressive diffusion index (ARDI) model of 
\cite{stock2002forecasting, stock2002macroeconomic}. Let $Z_{t}=\left[
y_{t},y_{t-1}...,y_{t-P_{y}},F_{t},F_{t-1}...,F_{t-P_{f}}\right] $ be our
feature matrix, then the ARDI model is given by 
\begin{eqnarray}
y_{t+h} &=&\beta Z_{t}+\epsilon _{t+h}  \label{ardi1} \\
X_{t} &=&\Lambda F_{t}+u_{t}  \label{ardi2}
\end{eqnarray}%
where $F_{t}$ are $k$ factors extracted by principal components from the $%
N_{X}$-dimensional set of predictors $X_{t}$ and parameters are estimated by
OLS. The AR model is obtained by keeping in $Z_{t}$ only the lagged values
of $y_{t}$. 
The hyperparameters of both models are specified using the Bayesian
information criterion (BIC).

\vskip 0.2cm

\noindent \textsc{\textbf{Ridge, Lasso, and Elastic Net}.} The Elastic Net
model simultaneously predicts the target variable $y_{t+h}$ and selects the
most relevant predictors from a set of $N_Z$ features contained in $Z_t$
whose weights $\beta := (\beta_i)_{i=1}^{N_Z}$ solve the following penalized
regression problem 
\begin{equation*}
\hat{\beta} := \text{arg} \underset{\beta}{\min} \sum_{t=1}^T \left( y_{t+h}
- Z_t \beta \right)^2 + \lambda \sum_{i=1}^{N_Z} \left( \alpha |\beta_i| +
(1-\alpha) \beta_i^2 \right)
\end{equation*}
and where $(\alpha, \lambda)$ are hyperparameters. Here, $Z_t$ contains
lagged values of $y_t$, factors and $X_t$. The Lasso estimator is obtained
when $\alpha=1$, while the Ridge estimator imposes $\alpha=0$ and both use
unit weights throughout. We select $\lambda$ and $\alpha$ with grid search
where $\alpha \in \{.01,.02,.03,...,1\}$ and $\lambda \in [0,\lambda_{max}]$
where $\lambda_{max}$ is the penalty term beyond which coefficients are
guaranteed to be all zero assuming $\alpha \neq 0$. Since those algorithms
performs shrinkage (and selection), we do not cross-validate $P_y$, $P_f$
and $k$. We impose $P_y = 6$, $P_f = 6$ and $k = 8$ and let the algorithms
select the most relevant features for forecasting task at hand.

\vskip 0.2cm

\noindent \textsc{\textbf{Random Forests}.} This algorithm provides a means
of approximating nonlinear functions by combining regression trees. Each
regression tree partitions the feature space defined by $Z_t$ into distinct
regions and, in its simplest form, uses the region-specific mean of the
target variable $y_{t+h}$ as the forecast, i.e. for $M$ leaf nodes 
\begin{align*}
\hat{y}_{t+h} = \sum_{m=1}^M c_m I_{(Z_t \in R_m)}
\end{align*}
where $R_1,...,R_M$ is a partition of the feature space. The input $Z_t$ is
the same as in the case of Elastic Net models. To circumvent some of the
limitations of regression trees, \cite{breiman2001} introduced Random
Forests. Random Forests consist in growing many trees on subsamples (or
nonparametric bootstrap samples) of observations. A random subset of
features is eligible for the splitting variable, further decorrelating them.
The final forecast is obtained by averaging over the forecasts of all trees.
In this paper we use 500 trees which is normally enough to stabilize the
predictions. The minimum number of observation in each terminal nodes is set
to 3 while the number of features considered at each split is $\frac{\# Z_t}{%
3}$. In addition, we impose $P_y = 6$, $P_f = 6$ and $k = 8$.

\vskip 0.2cm

\noindent \textsc{\textbf{Boosted Trees}.} This algorithm provides an
alternative means of approximating nonlinear functions by additively
combining regression trees in a sequential fashion. Let $\eta \in \lbrack
0,1]$ be the learning rate and $\hat{y}_{t+h}^{(n)}$ and $%
e_{t+h}^{(n)}:=y_{t-h}-\eta \hat{y}_{t+h}^{(n)}$ be the step $n$ predicted
value and pseudo-residuals, respectively. Then, for square loss, the step $%
n+1$ prediction is obtained as 
\begin{equation*}
\hat{y}_{t+h}^{(n+1)}=y_{t+h}^{(n)}+\rho _{n+1}f(Z_{t},c_{n+1})
\end{equation*}%
where $(c_{n+1},\rho _{n+1}):=\text{arg}\underset{\rho ,c}{\min }%
\sum_{t=1}^{T}\left( e_{t+h}^{(n)}-\rho _{n+1}f(Z_{t},c_{n+1})\right) ^{2}$
and $c_{n+1}:=\left( c_{n+1,m}\right) _{m=1}^{M}$ are the parameters of a
regression tree. In other words, it recursively fits trees on
pseudo-residuals. We consider a vanilla Boosted Trees where the maximum
depth of each tree is set to 10 and all features are considered at each
split. We select the number of steps and $\eta \in \lbrack 0,1]$ with
Bayesian optimization. $Z_t$ contains lagged values of $y_t$, factors and $%
X_t$, and we impose $P_{y}=6$, $P_{f}=6$ and $k=8$.

\vskip 0.2cm

\noindent \textsc{\textbf{Kernel Ridge Regressions}.} A way to introduce
high-order nonlinearities among predictors' set $Z_t$, but without
specifying a plethora of basis functions, is to opt for the Kernel trick. As
in \cite{GCLSS2019}, the nonlinear ARDI predictive equation (\ref{ardi1}) is
written in a general nonlinear form $g(Z_t)$ and can be approximated with
basis functions $\phi()$ such that 
\begin{equation*}
y_{t+h} = g(Z_t) + \varepsilon_{t+h} = \phi(Z_t)^{\prime }\gamma +
\varepsilon_{t+h}.
\end{equation*}
The so-called Kernel trick is the fact that there exist a reproducing kernel 
$K()$ such that 
\begin{equation*}
\hat{E}(y_{t+h}|Z_t) = \sum_{i=1}^t \hat{\alpha_i}\langle \phi(Z_i),
\phi(Z_t) \rangle = \sum_{i=1}^t \hat{\alpha_i}K( Z_i, Z_t ).
\end{equation*}
This means we do not need to specify the numerous basis functions, a
well-chosen kernel implicitly replicates them. Here we use the standard
radial basis function (RBF) kernel 
\begin{equation*}
{\displaystyle K_{\sigma}(\mathbf{x} ,\mathbf{x^{\prime }} )=\exp \left(-{%
\frac {\|\mathbf{x} -\mathbf{x^{\prime }} \|^{2}}{2\sigma ^{2}}}\right)}
\end{equation*}
where $\sigma$ is a tuning parameter to be chosen by cross-validation. In
terms of implementation, after factors are extracted via PCA from (\ref%
{ardi2}), the forecast of the Kernel Ridge Regression (KRR) diffusion index
model is obtained from 
\begin{equation*}
\hat{E}(y_{t+h}|Z_t)=K_{\sigma}(Z_t,Z)(K_{\sigma}(Z_t,Z)+\lambda
I_T)^{-1}y_t.  \label{KT4}
\end{equation*}
Here, we impose the same set of inputs, $Z_t$, as in the ARDI model and we
fix $P_{y}=6$, $P_{f}=6$ and $k=8$.

\vskip 0.2cm

\noindent \textsc{\textbf{Neural Networks}.} 
We consider standard feed-forward networks and the architecture closely
follows that of \cite{gu2018empirical}. Cross-validating the whole network
architecture is a difficult task especially with a small number of
observations as is the case in macroeconomic applications. Hence, we use two
hidden layers, the first with 32 neurons and the second with 16 neurons. The
number of epochs is fixed at 100. The activation function is ReLu and that
of the output layer is linear. The batch size is 32 and the optimizer is
Adam (Keras default values). The learning rate and the Lasso parameter are
chosen by 5-fold cross-validation among the following grids respectively, $%
\in \{0.001,0.01\}$ and $\in \{0.001,0.0001\}$. We apply the early stopping,
i.e. we wait for 20 epochs to pass without any improvement of the
cross-validation MSE to stop the training. The final prediction is the
average of an ensemble of 5 different estimations. $Z_t$ contains lagged
values of $y_t$, factors and $X_t$, and we impose $P_{y}=6$, $P_{f}=6$ and $%
k=8$.

\subsection{Additional Forecasting Models}

\noindent \textsc{\textbf{Macroeconomic Random Forests}.} \cite{MRF}
proposes a new form of RF better suited for macroeconomic data. The new
problem is to extract generalized time-varying parameters (GTVPs) 
\begin{align*}
y_{t}& =\tilde{X}_{t}\beta _{t}+\epsilon _{t} \\
\beta _{t}& =\mathcal{F}(S_{t})
\end{align*}%
where $S_{t}$ are the state variables governing time variation and $\mathcal{%
F}$ a forest. $S_{t}$ is (preferably) a high-dimensional macroeconomic data
set. In this paper, it is the same $Z_{t}$ as in plain RF and Boosting. $%
\tilde{X}$ determines the \textit{linear} model that we want to be
time-varying. Usually $\tilde{X}\subset S$ is rather small (and focused)
compared to $S$. For instance, an autoregressive random forests (ARRF) uses
lags of $y_{t}$ for $\tilde{X}_{t}$. A factor-augmented ARRF (FA-ARRF) adds
factors to ARRF's linear part.

The problem is to find the optimal variable $S_{j}$ (so, finding the best $j$
out of the random subset of predictors indexes $\mathcal{J}^{-}$) to split
the sample with, and at which value $c$ of that variable should we split.
The outputs should be $j^{\ast }$ and $c^{\ast }$ to be used to split $l$
(the parent node) into two children nodes, $l_{1}$ and $l_{2}$. Hence, the
greedy algorithm developed in \cite{MRF} runs 
\begin{equation}
\begin{aligned} \min _{j \in \mathcal{J^-}, \enskip c \in \rm I\!R} \Bigg[ &
\min _{\beta_{1}}\sum_{t \in l_1^{RW}(j,c)} w(t;\zeta)
\left(y_{t}-\tilde{X}_{t}\beta_{1}\right)^{2}+\lambda \Vert \beta_1 \Vert_2
\\ + & \min _{\beta_{2}} \sum_{t \in l_2^{RW}(j,c)}
w(t;\zeta)\left(y_{t}-\tilde{X}_{t} \beta_{2}\right)^{2}+\lambda \Vert
\beta_2 \Vert_2 \Bigg]. \end{aligned}  \label{mrf_algo_rw}
\end{equation}%
recursively to construct trees.

As it was the case for RF, the bulk of regularization comes from taking the
average over a diversified ensemble of trees (generated by both Bagging and
a random $\mathcal{J}^{-}\subset \mathcal{J}$. Nonetheless, $\beta_t$'s (and
the attached prediction) can also benefit from extra (yet mild)
regularization. Time-smoothness is made operational by taking the
"rolling-window view" of time-varying parameters. That is, the tree solve
many weighted least squares problems (WLS) which includes close-by
observations. To keep computational demand low, the kernel $w(t;\zeta)$ is a
symmetric 5-step Olympic podium. Informally, the kernel puts a weight of 1
on observation $t$, a weight of $\zeta<1$ for observations $t-1$ and $t+1$
and a weight of $\zeta^2$ for observations $t-2$ and $t+2$. Note that a
small Ridge penalty is added to make sure every matrix inverts nicely (even
in very small leaves), so a single tree has in fact two sources of
regularization. 

The standard RF is a restricted version of MRF where $\tilde{X}_{t}=\iota $, 
$\lambda =0$, $\zeta =0$ and the block size for Bagging is 1. In words, the
only regressor is a constant, there is no within-leaf shrinkage, and Bagging
does not care for serial dependence. It is understood that MRF will have an
edge over RF whenever linear signals included in $\tilde{X}_{t}$ are strong
and the number of training observations (or signal-to-noise ratio) is low.
The reason for this is simple: MRF nudge the learning algorithm in the right
direction rather than hoping for RF to learn everything non-parametrically.
Moreover, by providing generalized time-varying parameters (and credible
regions for those), MRF lends itself more easily to interpretation.

\vskip 0.2cm

\noindent \textsc{\textbf{Moving Average Rotation of $X$}.} The Moving
Average Rotation of $X$ (MARX) transformation was proposed in \cite{MDTM} as
a feature engineering technique which generates an implicit shrinkage more
appropriate for time series data. In linear setup when coefficients are
shrunk (and maybe selected) to 0, using MARX transform the usual $\beta
_{k,p}\rightarrow0$ prior into shrinking each $\beta _{k,p}$ to $\beta
_{k,p-1}$ for the $p$ lag of predictor $k$. For more sophisticated
techniques where shrinkage is only implicit (like RF and Boosting), MARX
"proposes" the variable-selecting algorithm with pre-assembled group of lags
which helps in avoiding that the underlying trees waste splits on a bunch of
scattered lags \citep{MRF}. \cite{MDTM} report that the transformation is
particularly helpful for US monthly real economic activity targets. Adding
MARX to the input set $Z_t$ is considered in all models except ARDI and
KRR.\medskip

\begin{table}[h]
\begin{center}
\begin{threeparttable}
			\caption{Forecasting Models}\label{fcst_models}
			\footnotesize
			\centering
			\rowcolors{2}{white}{gray!15}
			\setlength{\tabcolsep}{0.65em}
			\renewcommand{\arraystretch}{1.3}
			\begin{tabular}{lll}
				\toprule
				\toprule
				
				Name & Acronym   & \multicolumn{1}{l}{Input ($Z_t$)} \\ 
				\midrule
				Autoregression (with $P_y$ chosen by BIC) & \textbf{\textcolor{black}{AR,BIC}}   & $[\hspace*{0.1cm} y_{t-\{1:6\}}]$ \\
				Random Walk &  \textbf{\textcolor{black}{RW}} & $\emptyset$ \\
				Factor-Augmented AR (with $P_y$, $M_k$ and $K$ chosen by BIC) & \textbf{ARDI,BIC}   & $[\hspace*{0.1cm} y_{t-\{1:6\}}, \hspace*{0.1cm} F_{1:8,t-\{1:6\}}] $ \\
				\midrule
				LASSO & \textbf{LASSO}   & $[\hspace*{0.1cm} y_{t-\{1:6\}}, F_{1:8,t-\{1:6\}}, X]$ \\
				LASSO using MARX & \textbf{LASSO+MARX}   & $[\hspace*{0.1cm} y_{t-\{1:6\}}, F_{1:8,t-\{1:6\}}, X, $MARX$]$ \\
				Ridge & \textbf{RIDGE}   &  $[\hspace*{0.1cm} y_{t-\{1:6\}}, F_{1:8,t-\{1:6\}}, X]$  \\
				Ridge using MARX & \textbf{RIDGE+MARX}   & $[\hspace*{0.1cm} y_{t-\{1:6\}}, F_{1:8,t-\{1:6\}}, X, $MARX$]$ \\
				Elastic-Net & \textbf{E-NET}   & $[\hspace*{0.1cm} y_{t-\{1:6\}}, F_{1:8,t-\{1:6\}}, X]$ \\
				Elastic-Net using MARX & \textbf{E-NET+MARX} &  $[\hspace*{0.1cm} y_{t-\{1:6\}}, F_{1:8,t-\{1:6\}}, X, $MARX$]$ \\
				\midrule
				Kernel Ridge Regression & \textbf{KRR}   & $[\hspace*{0.1cm} y_{t-\{1:6\}}, F_{1:8,t-\{1:6\}}]$ \\
				Random Forest & \textbf{RF}   & $[\hspace*{0.1cm} y_{t-\{1:6\}}, F_{1:8,t-\{1:6\}}, X]$ \\
				Random Forest using MARX & \textbf{RF+MARX}   & $[\hspace*{0.1cm} y_{t-\{1:6\}}, F_{1:8,t-\{1:6\}}, X, $MARX$]$ \\
				Boosting & \textbf{Boosting}   & $[\hspace*{0.1cm} y_{t-\{1:6\}}, F_{1:8,t-\{1:6\}}, X]$ \\
				Boosting using MARX & \textbf{Boosting+MARX}   & $[\hspace*{0.1cm} y_{t-\{1:6\}}, F_{1:8,t-\{1:6\}}, X, $MARX$]$ \\
				AR Random Forest (linear part is $[\hspace*{0.1cm} y_{t-\{1:2\}}]$) & \textbf{ARRF(2)} & $[\hspace*{0.1cm} y_{t-\{1:6\}}, F_{1:8,t-\{1:6\}}, X, $MARX$]$ \\
				AR Random Forest (linear part is $[\hspace*{0.1cm} y_{t-\{1:6\}}]$) & \textbf{ARRF(6)} & $[\hspace*{0.1cm} y_{t-\{1:6\}}, F_{1:8,t-\{1:6\}}, X, $MARX$]$  \\
				Factor-Augmented AR RF (linear part is $[\hspace*{0.1cm} y_{t-\{1:2\}},  \hspace*{0.1cm} F_{1:2,t-1}]$) & \textbf{FA-ARRF(2,2)} & $[\hspace*{0.1cm} y_{t-\{1:6\}}, F_{1:8,t-\{1:6\}}, X, $MARX$]$    \\
				Factor-Augmented AR RF (linear part is $[\hspace*{0.1cm} y_{t-\{1:2\}},  \hspace*{0.1cm} F_{1:4,t-1}]$) & \textbf{FA-ARRF(2,4)} & $[\hspace*{0.1cm} y_{t-\{1:6\}}, F_{1:8,t-\{1:6\}}, X, $MARX$]$ \\
				Neural Network & \textbf{NN-ARDI} & $[\hspace*{0.1cm} y_{t-\{1:6\}}, F_{1:8,t-\{1:6\}}, X]$ \\
				Neural Network using MARX & \textbf{NN-ARDI+MARX} & $[\hspace*{0.1cm} y_{t-\{1:6\}}, F_{1:8,t-\{1:6\}}, X, $MARX$]$ \\
				\bottomrule
				\bottomrule
			\end{tabular}
			\begin{tablenotes}[para,flushleft]
				
			\end{tablenotes}
		\end{threeparttable}
\end{center}
\par
\label{tab:listmodels}
\end{table}

\section{UK-MD: A Large UK Monthly Macroeconomic Data Set}

\label{sec:UKMD}

Large datasets are now very popular in empirical macroeconomic research
since \cite{stock2002forecasting,stock2002macroeconomic} have initiated the
breakthrough by providing the econometric theory and showing the benefits in
terms of macroeconomic forecasting. \cite%
{mccracken2016fred,mccracken2020fred} proposed a standardized version of a
large monthly and quarterly US datasets that are regularly updated and
publicly available at the FRED (Federal Reserve Economic Data) website. \cite%
{fglss2018} have developed the Canadian version of FRED. In this paper, we
construct a similar large-scale UK macroeconomic database in monthly
frequency that can be used in the same way as the US and the Canadian data
sets. The dataset is described in the first subsection and analyzed in the
second one.

\subsection{UK-MD}

The dataset contains 112 macroeconomic and financial indicators divided into
nine categories: labour, production, retail and services, consumer and
retail price indices, producer price indices, international trade, money,
credit and interest rate, stock market and finally sentiment and leading
indicators. The selection of variables is inspired by \cite%
{mccracken2016fred}, \cite{fglss2018} and \cite{jkpk2021}. The complete list
of series is available in the data appendix \ref{sec:UKdataappendix}. Most
of the indicators are available at the Office of National Statistics, while
others are taken from the Bank of England, FRED and Yahoo finance. The
starting date varies across indicators, from 1960 to 2000. For the
forecasting application in this paper, data start in 1998M01.

Most of the series included in the database must be transformed to induce
stationarity. We roughly follow \cite{mccracken2016fred} and \cite{fglss2018}%
. For instance, most I(1) series are transformed in the first difference of
logarithms; a first difference of levels is applied to unemployment rate and
interest rates; and the first difference of logarithms is used for all price
indices. Transformation codes are reported in data appendix.

Our last concern is to balance the resulting panel since some series have
missing observations. We opted to apply an expectation-maximization
algorithm by assuming a factor model to fill in the blanks as in \cite%
{stock2002macroeconomic} and \cite{mccracken2016fred}. We initialize the
algorithm by replacing missing observations with their unconditional mean,
starting in 1998M1, and then proceed to estimate a factor model by principal
component. The fitted values of this model are used to replace missing
observations.

Finally, for this application we also add nineteen US macroeconomic and
financial aggregates as considered in \cite{bgr2008}. These series include
income, production, labour market, housing, consumption and monetary
indicators, as well as interest rates and prices. The complete list is
available in the appendix \ref{sec:USdataappendix}.

\subsection{Exploring the Factor Structure of UK-MD}

Large macroeconomic datasets are mainly used for forecasting and impulse
response analysis through lenses of factor modeling %
\citep{kotchoni2019macroeconomic,Bernanke-Boivin-Eliasz(2005)}. Indeed, the
factors provide a widely used dimension reduction method, but they also
serve as an empirical representation of general equilibrium models %
\citep{Boivin-Giannoni(2006)}. Hence, it is important to explore the factor
structure of our UK-MD dataset.

Estimating the number of factors is an empirical challenge and several
statistical decision procedures have been proposed, see \cite%
{Mao-Stevanovic(2015)} for review. Here, we select the number of static
factors using the \cite{Bai-Ng(2002)} $PC_{p2}$ criterion, and we follow 
\cite{Hallin-Liska(2007)} to test for the number of dynamic factors. $%
PC_{p2} $ criterion finds eight significant factors, while the number of
dynamic components is estimated at four. In addition, we performed the \cite%
{Alessi-Barigozzi-Capaso(2010)} improvement of the $PC_{p2}$ criterion that
in turn suggests nine factors.

After the static factors are estimated by principal components as in \cite%
{stock2002forecasting}, we report in Table \ref{tab:factorinterpretation}
their marginal contribution to the variance of variables constituting UK-MD.
For instance, $mR^2_i(k)$ measures the incremental explanatory power of the
factor $k$ for the variable $i$, which is simply the difference between the $%
R^2$ after regressing the variable $i$ on the first $k$ and $k-1$ factors.
The overall marginal contribution of the factor $k$ is the sample average
over all variables. Table \ref{tab:factorinterpretation} shows the average $%
mR^2(k)$ for each of nine estimated factors, lists ten series that load most
importantly on each factor and indicates the group to which the series
belongs. For example, factor 1 explains 20.7\% of the variation in UK-MD and
is clearly a real activity factor as the ten most related variables are
indicators of production and services. In particular, it explains 88.7 and
83.6\% of variation in the index of services and the index of production in
manufacturing respectively. The second factor explains 8.4\% of variation
overall, and represents mainly the group of interest rates. For instance,
its marginal contribution to the 12-month LIBOR is 0.532. Factor 3's average
explanatory power is 5.4\% and it is linked to prices indices, with the
highest $mR^2_i(k) = 0.513$ for the CPI inflation. Factors 4 and 5 are
related to stock market and employment variables respectively. The sixth
factor explain 3.4\% of total variation and can be interpreted as the
international trade factor. Factor 7 is related to unemployment and working
hours indicators, with an explanatory power of 24.5\% for the over 12 month
unemployment duration. Exchange rates are well explained by the seventh
factor. Finally, the ninth component stands out as an energy factor as it
explains a sizeable fraction of variation in production indices of oil
extraction, mining and energy sectors.

\begin{table}[t]
\caption{Interpretation of factors estimated from UK-MD, 1998M1-2020M9 }
\label{tab:factorinterpretation}
\begin{center}
{\scriptsize 
\begin{tabular}{lll|lll|lll}
\toprule $mR^2(1)$ & 0,207 & G\# & $mR^2(2)$ & 0,084 & G\# & $mR^2(3)$ & 
0,054 & G\# \\ \hline
IOS & 0,887 & 3 & LIBOR\_12mth & 0,532 & 6 & CPI\_ALL & 0,513 & 4 \\ 
IOP\_MANU & 0,836 & 2 & LIBOR\_3mth & 0,486 & 6 & CPIH\_ALL & 0,466 & 4 \\ 
AVGW\_RET\_SALE\_NF & 0,810 & 3 & RPI\_ALL & 0,469 & 4 & CPI\_EX\_ENER & 
0,392 & 4 \\ 
IOP\_PROD & 0,802 & 2 & LIBOR\_1mth & 0,418 & 6 & CPI\_GOOD & 0,391 & 4 \\ 
IOS\_PNDS & 0,786 & 3 & BANK\_RATE & 0,411 & 6 & RPI\_GOOD & 0,238 & 4 \\ 
CLI & 0,781 & 8 & BGS\_5yrs\_yld & 0,366 & 6 & PPI\_MANU & 0,185 & 9 \\ 
IOP\_INT\_GOOD & 0,770 & 2 & RPI\_GOOD & 0,308 & 4 & RPI\_ALL & 0,182 & 4 \\ 
IOS\_45 & 0,768 & 3 & BGS\_10yrs\_yld & 0,287 & 6 & EMP\_RATE & 0,171 & 1 \\ 
IOS\_G & 0,765 & 3 & PPI\_MANU & 0,284 & 9 & RPI\_SERV & 0,171 & 4 \\ 
IOP\_CAP\_GOOD & 0,765 & 2 & MORT\_FRATE\_2YRS & 0,269 & 6 & CPI\_TRANS & 
0,169 & 4 \\ 
\midrule $mR^2(4)$ & 0,045 & G\# & $mR^2(5)$ & 0,038 & G\# & $mR^2(6)$ & 
0,034 & G\# \\ \hline
FTSE250 & 0,432 & 7 & EMP & 0,257 & 1 & EXP\_GOOD & 0,338 & 5 \\ 
FTSE\_ALL & 0,386 & 7 & EMP\_ACT\_RATE & 0,209 & 1 & EXP\_TOT & 0,290 & 5 \\ 
SP500 & 0,385 & 7 & EMP\_RATE & 0,197 & 1 & IMP\_GOOD & 0,197 & 5 \\ 
UK\_focused\_equity & 0,360 & 7 & EMP\_ACT & 0,188 & 1 & IMP\_FUEL & 0,188 & 
5 \\ 
EMP & 0,245 & 1 & FTSE\_ALL & 0,177 & 7 & EXP\_FUEL & 0,175 & 5 \\ 
EMP\_RATE & 0,210 & 1 & FTSE250 & 0,175 & 7 & IMP\_ALL & 0,160 & 5 \\ 
EUR\_UNC\_INDEX & 0,159 & 7 & UK\_focused\_equity & 0,144 & 7 & EXP\_MACH & 
0,153 & 5 \\ 
EMP\_PART & 0,152 & 1 & M4 & 0,142 & 6 & IMP\_OIL & 0,143 & 5 \\ 
EMP\_ACT & 0,152 & 1 & MORT\_FRATE\_2YRS & 0,138 & 6 & EXP\_OIL & 0,133 & 5
\\ 
EMP\_ACT\_RATE & 0,131 & 1 & LIBOR\_12mth & 0,128 & 6 & IMP\_MACH & 0,111 & 5
\\ 
\midrule $mR^2(7)$ & 0,033 & G\# & $mR^2(8)$ & 0,032 & G\# & $mR^2(9)$ & 
0,027 & G\# \\ \hline
UNEMP\_DURA\_12mth & 0,245 & 1 & GBP\_CAN & 0,277 & 5 & IOP\_OIL\_EXTRACT & 
0,530 & 2 \\ 
AVG\_WEEK\_HRS\_FULL & 0,186 & 1 & GBP\_BROAD & 0,264 & 5 & IOP\_MINE & 0,522
& 2 \\ 
AVG\_WEEK\_HRS & 0,185 & 1 & GBP\_EUR & 0,222 & 5 & IOP\_ENER & 0,469 & 2 \\ 
TOT\_WEEK\_HRS & 0,132 & 1 & EXP\_FUEL & 0,125 & 5 & EXP\_OIL & 0,138 & 5 \\ 
EMP\_RATE & 0,132 & 1 & M1 & 0,120 & 6 & EXP\_FUEL & 0,101 & 5 \\ 
UNEMP\_DURA\_24mth & 0,130 & 1 & PPI\_MACH & 0,111 & 9 & IMP\_CRUDE\_MAT & 
0,089 & 5 \\ 
UNEMP\_RATE & 0,128 & 1 & FTSE\_ALL & 0,111 & 7 & IMP\_METAL & 0,088 & 5 \\ 
AWE\_PRIV & 0,124 & 1 & EXP\_OIL & 0,108 & 5 & EXP\_MACH & 0,064 & 5 \\ 
VAC\_TOT & 0,124 & 1 & PPI\_MOTOR & 0,095 & 9 & EXP\_CRUDE\_MAT & 0,050 & 5
\\ 
AWE\_ALL & 0,109 & 1 & SP500 & 0,095 & 7 & EXP\_METAL & 0,043 & 5 \\ 
\bottomrule &  &  &  &  &  &  &  & 
\end{tabular}
}
\end{center}
\par
\flushleft
{\scriptsize \vspace{-2em} {\singlespacing
Note: This table shows the ten series that load most importantly on the
first nine factors. For example, the first factor explains 20.7\% of the
variation in all 112 series, and it explains 88.7\% of variation in IOS
indicator. The third column of each panel indicates the group to which the
variable belongs. Group 1: labour market. Group 2: production. Group 3:
retail and services. Group 4: consumer and retail price indices. Group 5:
international trade. Group 6: money, credit and interest rates. Group 7:
stock market. Group 8: sentiment and leading indicators. Group 9: producer
price indices. }}
\par
{\scriptsize \ }
\end{table}

\begin{figure}[t!]
\caption{Importance of factors}
\label{fig:factorimportance}\centering
\includegraphics[width=\textwidth,height=0.5%
\textheight]{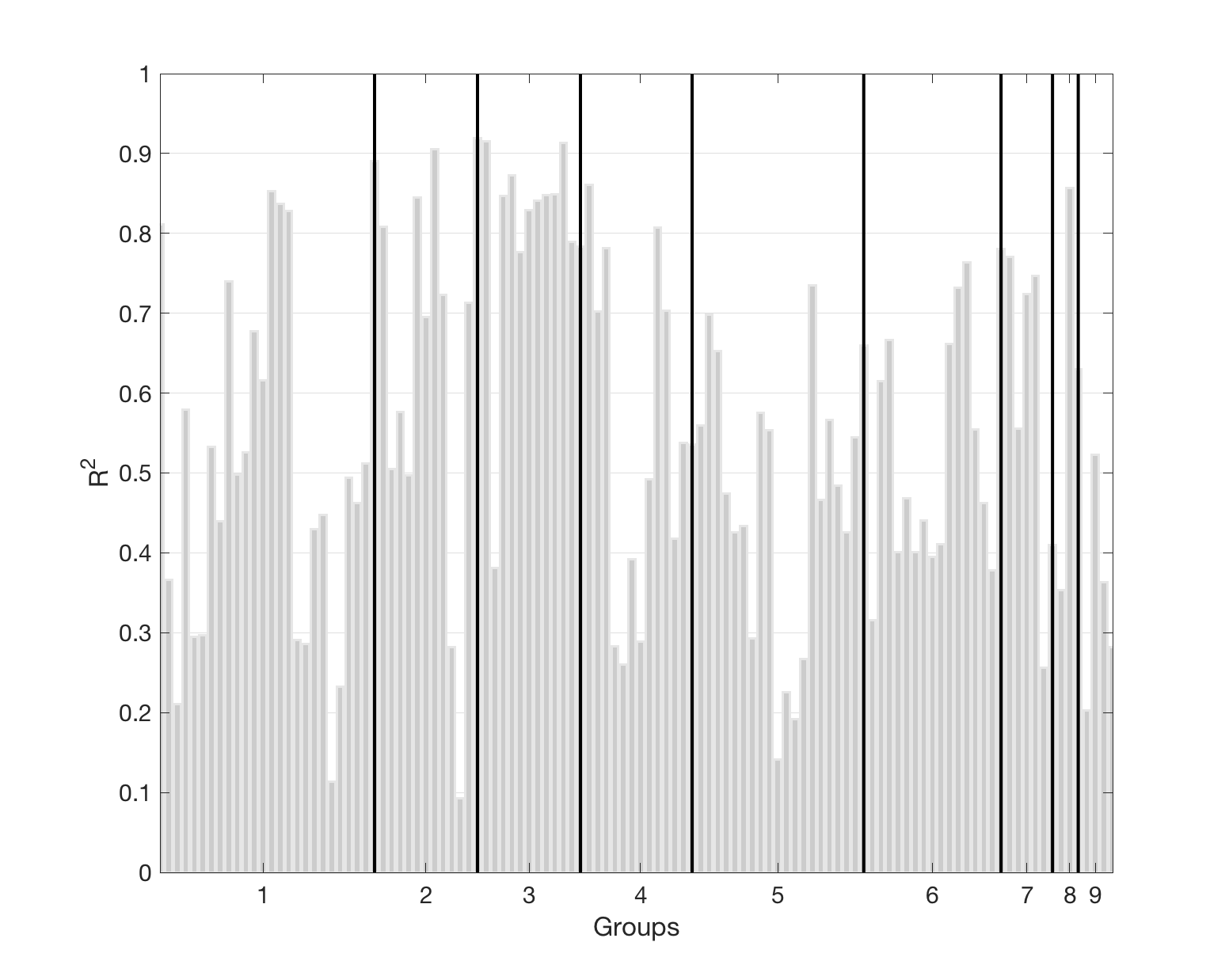} \flushleft
{\scriptsize \ \vspace{-3em} {\singlespacing
Note: This figure illustrates the explanatory power of the first nine
factors in the UK-MD series organized into nine groups. Group 1: labour
market. Group 2: production. Group 3: retail and services. Group 4: consumer
and retail price indices. Group 5: international trade. Group 6: money,
credit and interest rates. Group 7: stock market. Group 8: sentiment and
leading indicators. Group 9: producer price indices. }}
\par
{\scriptsize \ }
\end{figure}

Figure \ref{fig:factorimportance} plots the importance of the common
component with nine factors. The total $R^2$ is 0.554. The explanatory power
of the common component varies across series. It explains more than 80\% for
20 series, mostly services, production and average week hours series. The
nine factors are also very important for 42 variables as they have an $R^2$
between 0.5 and 0.8. There is only one series that have the idiosyncratic
component explaining over 90\% of the variation, IOP\_PETRO, and 3 variables
for which the common component $R^2$ is less than 20\%. Therefore, we can
conclude that the factor structure in UK-MD seems reasonable and is
comparable to those in FRED-MD and CAN-MD datasets. Interestingly, the
interpretation of the first three UK-MD factors is identical to the
interpretation of the first three FRED-MD components.

\begin{figure}[t!]
\caption{Number of factors and $R^2$ over time}
\label{fig:numberfactors}\centering
\includegraphics[width=\textwidth,height=0.5%
\textheight]{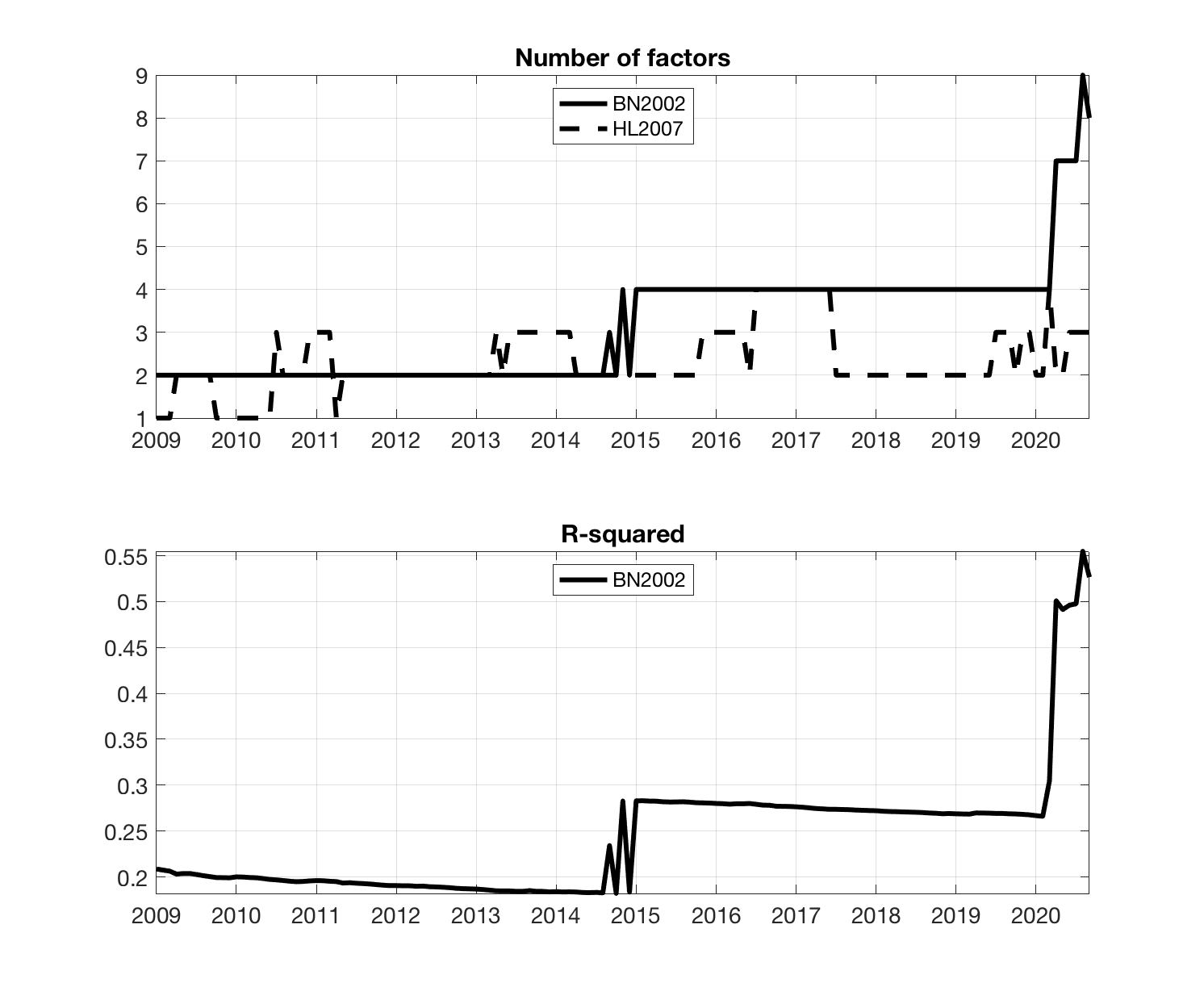} 
\flushleft
{\scriptsize \ \vspace{-3em} {\singlespacing
Note: This figure plots the number of factors selected recursively since
2009 by the \cite{Bai-Ng(2002)} $PC_{p2}$ criterion (upper panel) and the
corresponding total $R^2$ (bottom panel). }}
\par
{\scriptsize \ }
\end{figure}

In Figure \ref{fig:numberfactors} we show the number of static factors
selected recursively from 2009 by the \cite{Bai-Ng(2002)} $PC_{p2}$
criterion (upper panel) and the corresponding $R^{2}$ (bottom panel). The
number of significant factors increases over time. It goes from 
2 between 2009 and 2015, followed by a second plateau at 4 until 2020, and
it jumps to 7, 9 and 8 since the Covid-19 pandemic. The additional factors
emerging during the pandemic period are likely capturing the specificities
of this period.

\section{Empirical Setup}

\label{sec:empirics}

\subsection{Variables of Interest}

We focus on predicting twelve representative macroeconomic indicators of the
UK economy: Employment (EMP), Unemployment rate (UNEMP RATE), Total actual
weekly hours worked (HOURS), Industrial Production (IP PROD), Index of
production: manufacture of machinery and equipment (IP MACH), Total retail
trade (RETAIL), Consumer price index (CPI), Retail price index (RPI), RPI
Housing (RPI HOUSING), Consumer credit excluding student loans (CREDIT),
Total sterling approvals for house purchases (HOUSE APP) and Producer price
index of manufacturing sector (PPI MANU).

We consider the \emph{direct} predictive modeling in which the target is
projected on the information set, and the forecast is made directly using
the most recent observables. All the variables above are assumed $I(1)$, so
we forecast the average growth rate \citep{stock2002macroeconomic}, 
\begin{equation}
y_{t+h}^{(h)}=(1/h)\mathrm{ln}(Y_{t+h}/Y_{t}),  \label{fcst1}
\end{equation}%
except for UNRATE where we target the average change as in (\ref{fcst1}) but
without logs.

\subsection{Pseudo-Out-of-Sample Experiment Design}

The pseudo-out-of-sample period starts on 2008M01. The end period depends on
target variables. Labor market series, EMP, UNEMP RATE and HOURS, end on
2020M09, while RETAIL is available up to 2020M10. The rest of variables end
on 2020M11. The forecasting horizons considered are 1, 2 and 3 months. All
models are estimated recursively with an expanding window in order to
include more data so as to potentially reduce the variance of more flexible
models.

The standard \cite{dieboldmariano} (DM) test procedure is used to compare
the predictive accuracy of each model against the reference autoregressive
model. Mean squared error (MSE) is the most natural loss function given that
all models are trained to minimize the squared loss in-sample. 
Hyperparameter selection is performed using the BIC for AR and ARDI and
K-fold cross-validation is used for the remaining models. This approach is
theoretically justified in time series models under conditions spelled out
by \cite{bergmeir2018note}. Moreover, \cite{GCLSS2019} compared it with a
scheme which respects the time structure of the data in the context of
macroeconomic forecasting and found K-fold to be performing as well as or
better than this alternative scheme. All models are estimated (and their
hyperparameters re-optimized) every month. 

\section{Results}

\label{sec:results}

In this section we present the results of the forecasting experiment,
focusing first on the Covid-19 era and then on average performance over the
longer evaluation sample.

\subsection{Pandemic Recession Case Study}

\begin{figure}[t!]
\par
\begin{center}
\begin{subfigure}[b]{\textwidth}
  \begin{center}
\includegraphics[trim={1.5cm 0cm 0.25cm 0cm},clip,scale=.227]{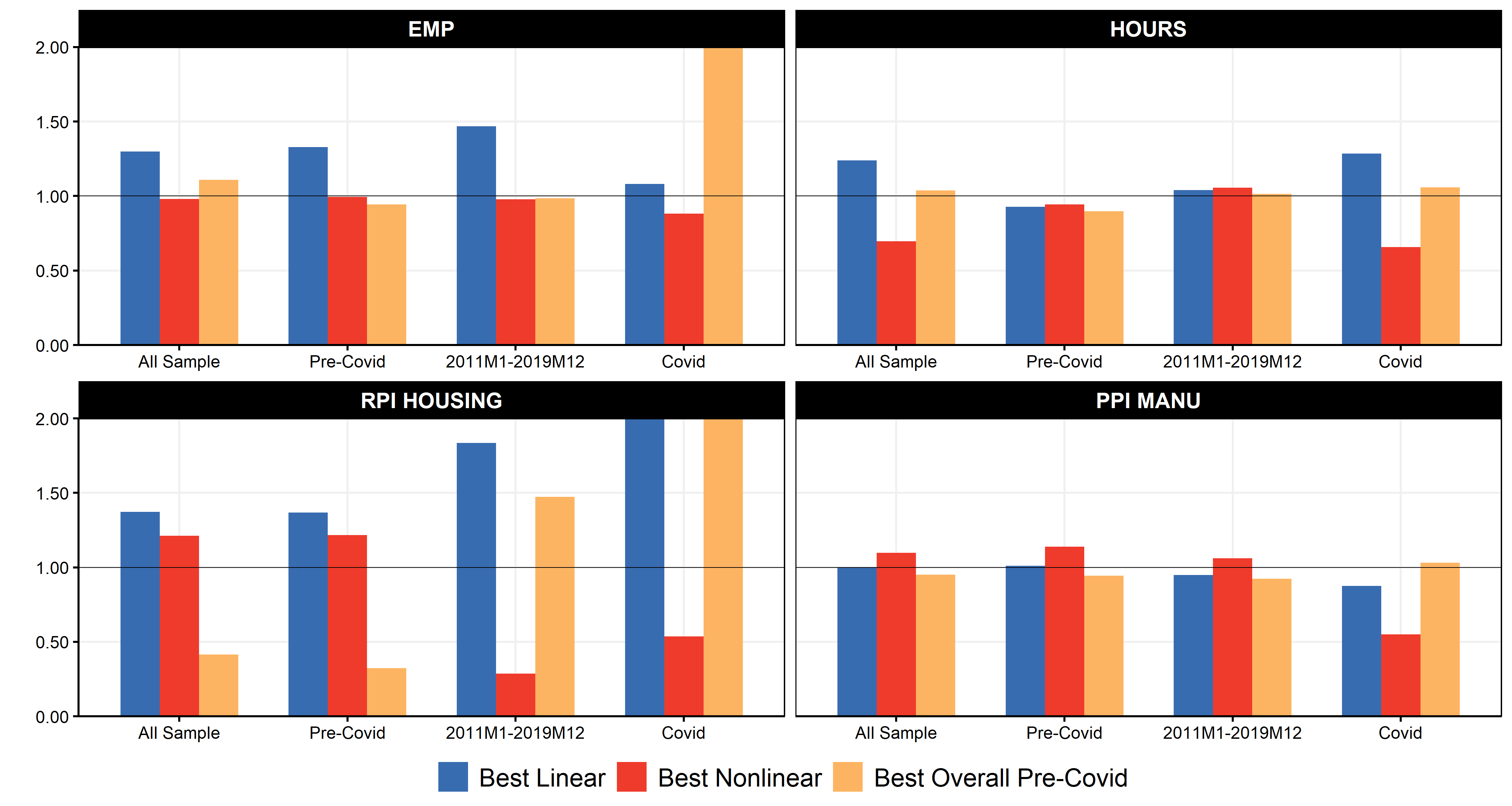}
\caption{MSEs wrt AR($p$)} 
\label{bigfour_mses} 
\end{center}
  \end{subfigure}
\begin{subfigure}[b]{\textwidth}
    \begin{center}
\includegraphics[trim={1.5cm 0cm 0.25cm 0cm},clip,scale=.295]{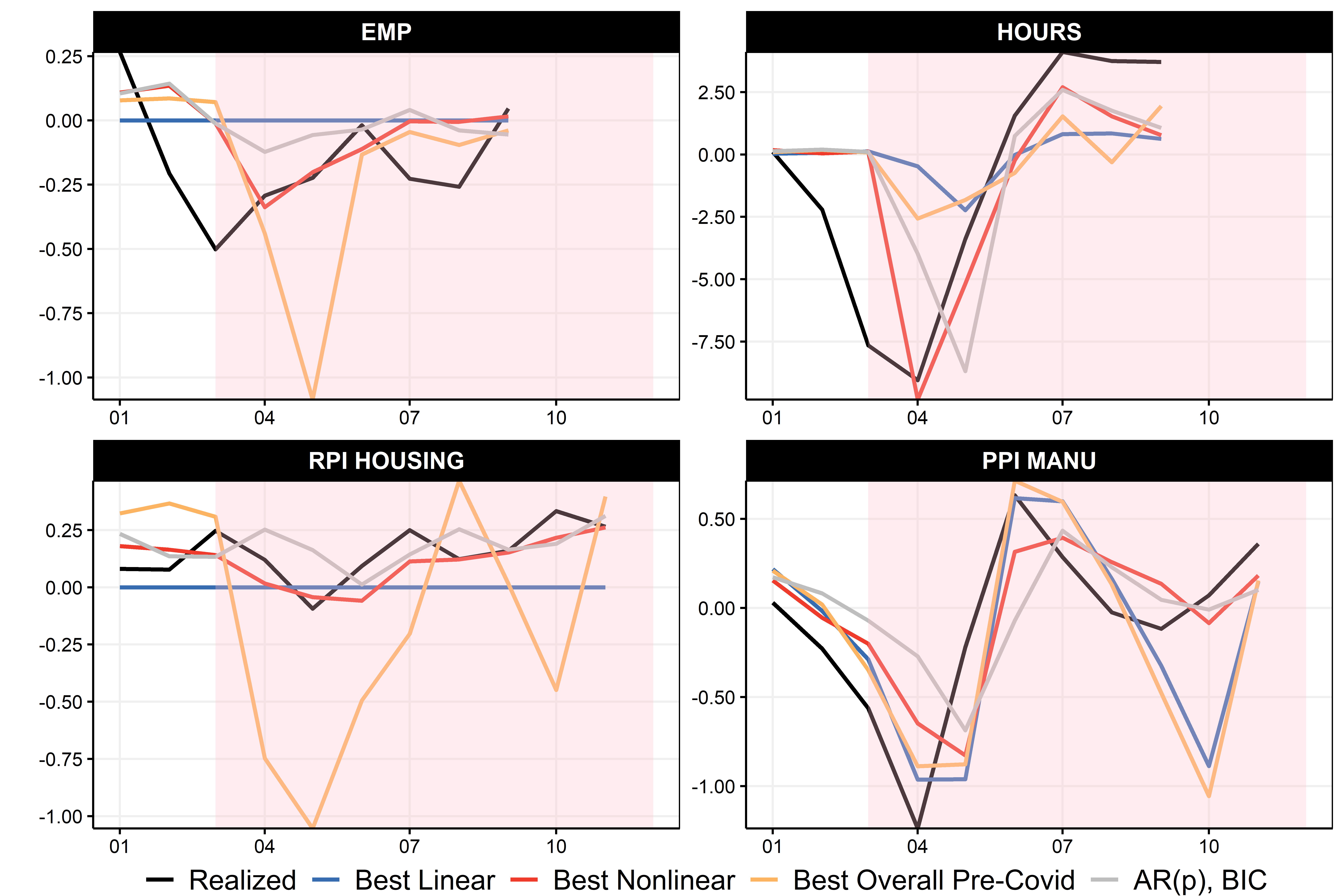}
\caption{Forecasts from January 2020}  
\label{bigfour_fcst}
\end{center}
  \end{subfigure}
\vspace*{-0.5cm}
\end{center}
\par
\vspace*{-1cm} \clearpage
\caption{Best Models for Four Selected Targets}
\label{bigfour}
\end{figure}

Figure \ref{bigfour} looks at four selected cases and compares the behavior
of the best models among certain categories: best linear model for the Covid
era, defined as the period 2020M1-2020M9/M11 depending on the variable, best
nonlinear model for the Covid era, and best model overall for the 2008-2019
period. The exact identities of selected models in Figure \ref{bigfour} are
reported in Table \ref{quad}.

\begin{table}[tbp]
\caption{Best COVID era Models (as displayed in Figure \protect\ref{bigfour}%
) }
\label{quad}
\vspace{-0.75cm} \hspace{-1.25cm} \setstretch{1.5} \setlength{%
\arrayrulewidth}{0.99pt}
\par
\begin{center}
{\footnotesize 
\begin{tabular}{l|l|c|c|c|c|}
\multicolumn{2}{c}{} & \multicolumn{4}{c}{\textbf{Variables}}\\
\hhline{~|~|-|-|-|-|}
\multicolumn{2}{c|}{}& \cellcolor{PineGreen!15}EMP & \cellcolor{PineGreen!15}HOURS & \cellcolor{PineGreen!15}RPI HOUSING & \cellcolor{PineGreen!15}PPI MANU \\
\hhline{~|-|-|-|-|-|}
\multirow{3}{*}{\textbf{Models}}& \cellcolor{PineGreen!15}Best Linear & RW  & RIDGE+MARX & RW & E-NET+MARX \\
\hhline{~|-|-|-|-|-|}
 &  \cellcolor{PineGreen!15}Best Nonlinear & FA-ARRF, 2Fac  & FA-ARRF, 4Fac & ARRF, 6Ylag & RF+MARX  \\
\hhline{~|-|-|-|-|-|}
 &  \cellcolor{PineGreen!15}Best Overall Pre-Covid  & RIDGE+MARX  & NN-ARDI  & LASSO+MARX & E-NET  \\
\hhline{~|-|-|-|-|-|}
\end{tabular}
}
\end{center}
\par
{\footnotesize \vspace*{-0.35cm} }
\end{table}

Though the Covid era is short and so the results should be interpreted with
care, the outcome is quite interesting. Linear models have a hard time
characterizing the path of \textbf{EMP} during the Pandemic recession.
Ridge+MARX, which was marginally better than the nonlinear FA-ARRF(2,2)
during the pre-Covid era, is predicting an employment cataclysm that did not
materialize. This is a general property of linear models for this target
since the best linear forecast (other than the AR) for EMP in 2020 is the 0
forecast, that is, the RW without drift in levels. FA-ARRF(2,4) (and
FA-ARRF(2,2) close behind) is the best model for EMP at a horizon of one
month. At longer horizons, RF-MARX is the best model, with a decisive
advantage over both AR and RF that do not use the transformations of \cite%
{MDTM}. This winning streak extends to unemployment at all horizons --
another variable that responded in a rather mild fashion to the Covid shock
due to Government intervention. Given RF usual robustness \citep{MSoRF},
those gains are almost all statistically significant.

\begin{figure}[h!]
\caption{GTVPs of FA-ARRF(2,2) --- EMP at $h=1$}
\label{fig:EMP_GTVP}\centering
\hspace*{-1.25em} \includegraphics[width=%
\textwidth]{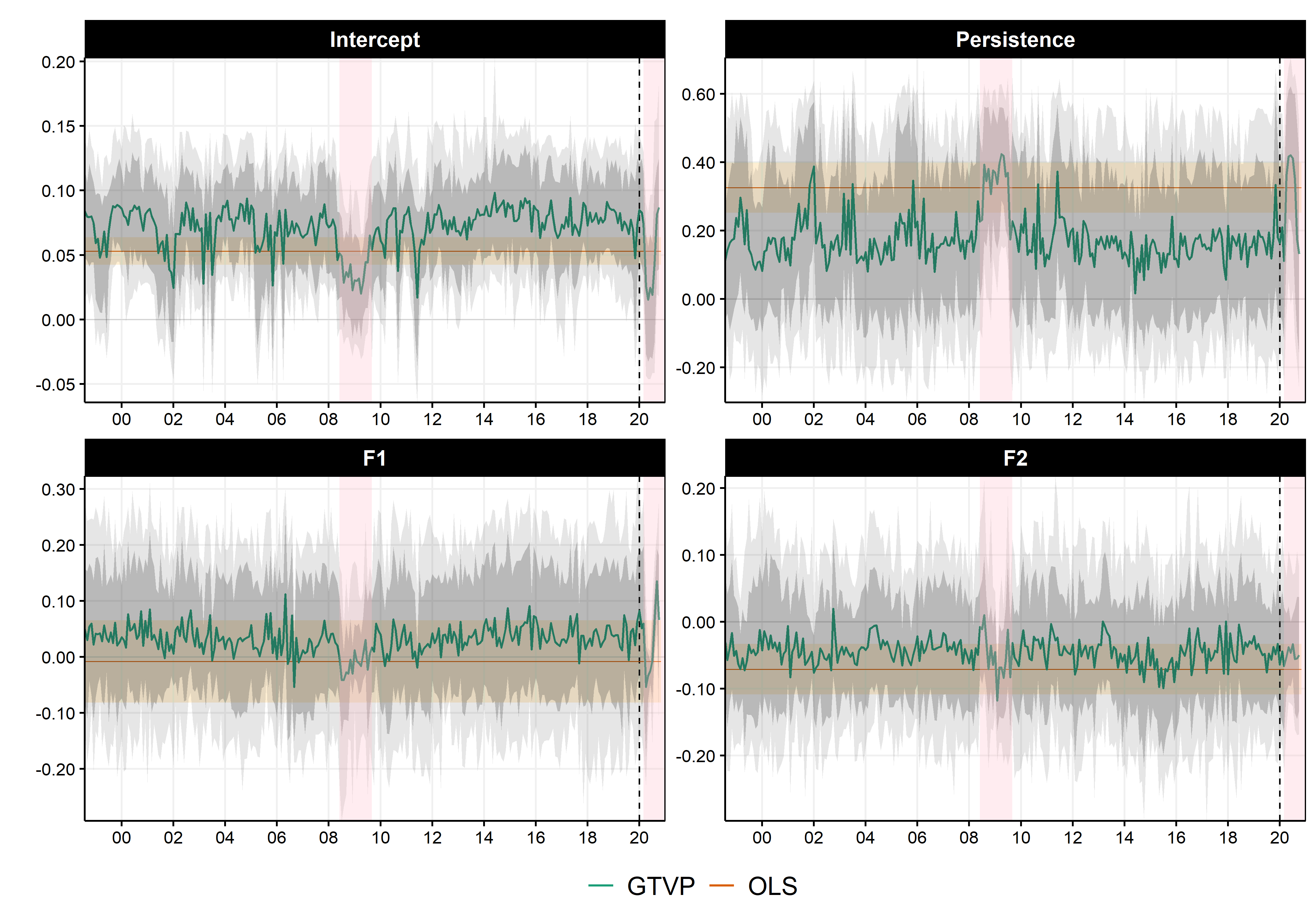}
\flushleft
{\scriptsize \ \vspace{-2em} {\singlespacing
Notes: GTVPs of the one month ahead EMP forecast. Persistence is defined as
the sum of $y_{t-1:2}$'s coefficients. The gray bands are the 68\% and 90\%
credible region. The pale orange region is the OLS coefficient $\pm$ one
standard error. The vertical dotted line is the end of the training sample
(for this graph only, not the forecasting exercise itself, which is
ever-updating). Pink shading corresponds to recessions. }}
\par
{\scriptsize \ }
\end{figure}

In Figure \ref{bigfour_fcst}, we see that the improvement at $h=1$ comes
from responding more swiftly (and more vigorously) to the first Covid shock
than what AR would allow for. An explanation for this well-calibrated
response can be found in Figure \ref{fig:EMP_GTVP} which plots the
underlying Generalized Time-Varying Parameters (GTVPs) for FA-ARRF(2,2). The
persistence seems to be highly state-dependent --- being much higher during
certain episodes (including recessions). This feature is replicated
out-of-sample during the Pandemic recession, which procured FA-ARRF(2,2) an
edge over the competitive plain AR. Additionally, the model incorporates an
intercept that alternates between two regimes, with the negative one being
attributed to recessions (but not exclusively according to pre-2008 data).
The drop in intercept is also predicted out-of-sample for the Covid period.
Unsurprisingly, those switches match those of persistence. Finally, it is
noted that the sensitivity to the first factor (which usually characterizes
real activity) is initially milder during recessions for EMP. This is a
salient feature for 2020 as the EMP response to the Covid shock is much
milder than that of other labor/production indicators (like HOURS).

Turning to \textbf{HOURS} -- which experienced an unprecedented rise and
fall during the onset of the Pandemic Recession --, it is striking to see
that only Macroeconomic Random Forests (MRF) can beat the AR benchmark at $%
h=1$. Indeed, the four MRFs report MSE ratios between 0.69 and 0.78 whereas
that of the other nonlinear models range between 1.05 and 1.5. Things are
even worse for linear models.

Figure \ref{fig:VI_hours} reports various variable importance (VI) measures
for FA-ARRF(2,2) (the reader is referred to \cite{MRF} for numerous
implementation details). Universally, the VIs suggest the predominance of
other labor indicators like measures of vacancies. Given how those are
closely related to HOURS itself, and that all successful MRFs include an AR
component, this points in the direction that HOURS may well follow a
nonlinear AR process which MRF is particularly well equipped to extract. As
a result, the response of MRF to the Covid shock is (as it was the case for
EMP), more timely than that of AR. Given how fast things were evolving back
in the spring of 2020, that timing provides MRF with an improvement of
around 30\% over the benchmark.

As conjectured earlier, MRF's capacity to extrapolate (which RF and Boosted
Trees both lack) proves vital for variables which exhibited (previously
unseen) swings of extraordinary proportions. While NN-ARDI also has the
capacity to extrapolate (and is marginally better than FA-ARRF(2,2) in the
pre-Covid era), its lack of an explicit linear part is likely to blame for
its spectacular incapacity to propel the Covid shock in Figure \ref%
{bigfour_fcst}. A similar dismal predicament is observed for RIDGE-MARX
which is the best linear model for the Covid sample.

Different troubles afflict data-rich linear models for \textbf{RPI HOUSING}
with MSE ratios exploding well over 10. As a result, the best linear model
is without question the simple autoregression. An obvious explanation for
the generalized failure of linear models (and also most data-rich ones) can
be found in Figure \ref{bigfour_fcst}. The "orange" forecasts basically
predict a path largely inspired by the experience of the Great Recession,
i.e., a joint collapse of real activity \textit{and} housing prices. Since
this is the sole recession in the training set, it is fair to say that most
ML methods naively (yet inevitably) associate real activity slowdown with a
significant drop in RPI Housing. However, by information available to the
economist, but not to the sample-constrained ML algorithm, this association
is more of a 2008-2009 exception than a "rule".

\begin{figure}[h!]
\caption{GTVPs of ARRF(6) --- RPI HOUSE at $h=1$}
\label{fig:GTVP_RPIhouse}\centering
\hspace*{-1.25em} \includegraphics[width=%
\textwidth]{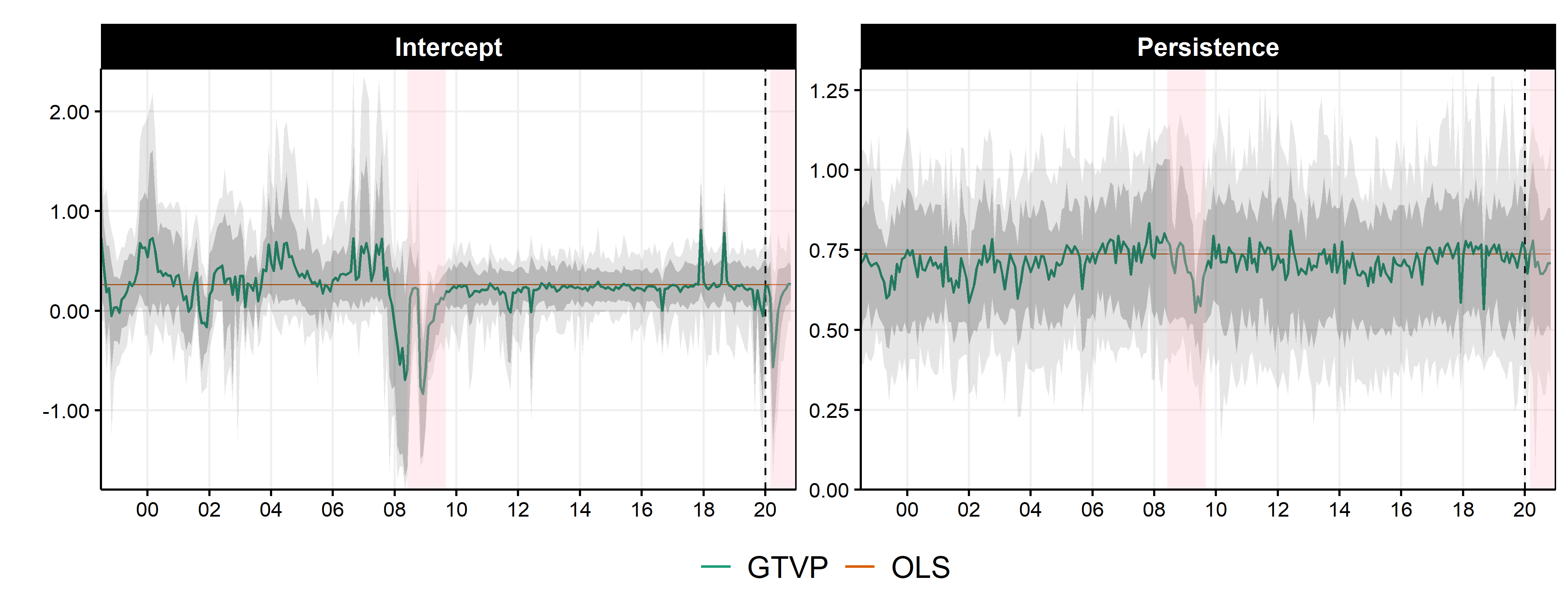} 
\flushleft
{\scriptsize \ \vspace{-2em} {\singlespacing
Notes: GTVPs of the one month ahead EMP forecast. Persistence is defined as
the sum of $y_{t-1:6}$'s coefficients. The reported intercept is the
long-run mean. The gray bands are the 68\% and 90\% credible region. The
pale orange region is the OLS coefficient $\pm$ one standard error. The
vertical dotted line is the end of the training sample (for this graph only,
not the forecasting exercise itself, which is ever-updating). Pink shading
corresponds to recessions. }}
\par
{\scriptsize \ }
\end{figure}

The only models able to beat the benchmark are the MRFs equipped with small
autoregressions as linear parts (ARRF(2) and ARRF(6)). So, how did they
avoid the dismal fates of other ML methods, and captured nicely the soft
drop (and bounce back) of RPI HOUSING in 2020? First, they do not rely
explicitly on linkage with other groups of variables (like FA-ARRFs would
through the use of factors) but rather focus on nonlinear autoregressive
dynamics. This strategy is expected to pay off whenever a shock can truly be
thought of as "exogenous" and we simply need a model to propagate it ---
this description corresponds to the onset of the Pandemic Recession but
certainly not its predecessor. Second, the model needs to separate pre-2008
dynamics from what followed. Figure \ref{fig:GTVP_RPIhouse} report
interesting transformations of ARRF(6)'s GTVPs. While persistence is rather
stable at 0.75, the long-run mean is subject to a lot of variation. Some is
cyclical (like the mild drops in 2008 and 2020), but the most noticeable
feature is a permanent regime change after 2008. Variable importance
measures in Figure \ref{fig:VI_RPIhouse} validate this observation: much of
the forest generating the time-variation uses either "trend" (i.e.,
exogenous change) or a catalog of indicators related to the policy rate (UK
Bank Rate, US Federal Funding Rate, and many MARX transformations of those)
whose are known to have entered uncharted territory in the aftermath of the
2008-2009 recession. Figure \ref{fig:forecasts_rpi} confirms visually that
the variation in the intercept of ARRF(6) gives an edge over both AR and the
best linear model (RIDGE-MARX), especially starting from 2011. As a result,
ARRF(6) is also the best model for all horizons in the quieter period of
2011-2019 (see Table \ref{tab:quiet}) with improvements over the AR
benchmark of 70\%, 54\% and 54\% at horizons 1 to 3 respectively.

\begin{figure}[h!]
\caption{Variable Importance for RF and Boosting --- PPI MANU at $h=1$}
\label{fig:VI_ppimanu}\centering
\hspace*{-2em} %
\includegraphics[scale=0.45]{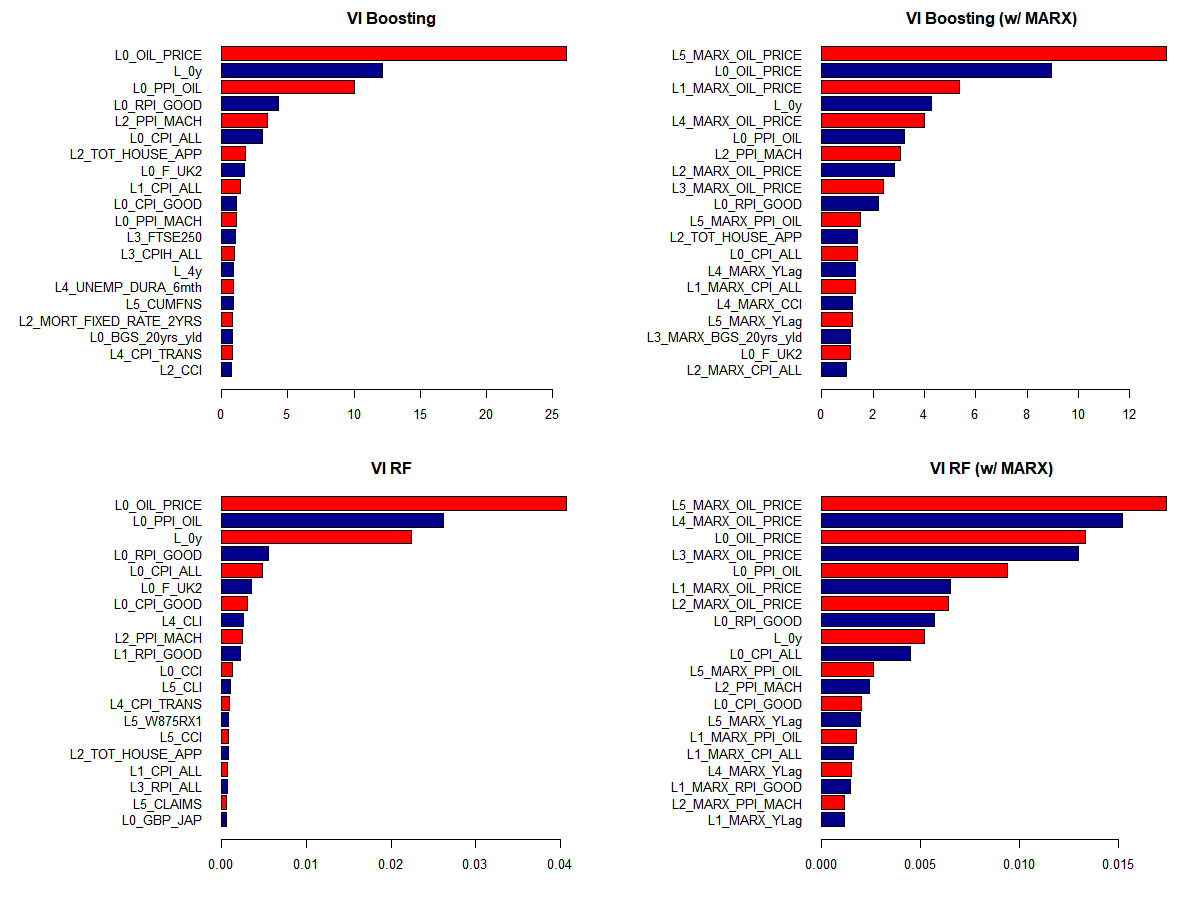} 
\flushleft
{\scriptsize \ \vspace{-2em} {\singlespacing
Note: Comparing Variable Importance for Boosting and RF, with and without
MARX, when forecasting PPI MANU at a one-month horizon. }}
\par
{\scriptsize \ }
\end{figure}

The last quadrant of Figure \ref{bigfour_mses} shows that for \textbf{PPI
MANU}, a model that does marginally worse most of the time can generate
substantial gain during the Covid period. Such is the case for RF-MARX which
performance is similar to that of the best linear model for most samples
(and the best overall pre-Covid). Figure \ref{bigfour_fcst} makes clear that
this edge during the Pandemic happens because (i) RF-MARX goes almost as
deep as linear models during the spring and yet (ii) does not call for a
large decrease in September and October (unlike linear models, and akin to
AR's prediction). Since RF-MARX does better than plain RF by 36\% and
Boosting-MARX better than plain Boosting by 12\%, it is natural curiosity to
investigate the VI measures of those models to uncover what particular MARX
transformations RF is so fond of. In Figure \ref{fig:VI_ppimanu}, we see
that both plain Boosting and RF rely strongly on the most recent values of
oil prices, PPI oil and PPI MANU itself --- which comes to no surprise.
Interestingly, the other lags of oil prices are generally absent from the
top 20. The MARX versions consider a slightly less focused set of predictors
composed of various moving averages of oil prices. In both the RF and
Boosting case, the most important feature is the last 6 months average of
oil prices change. Thus, RF-MARX versions avoid calling for another decrease
of PPI MANU by relying less on monthly oil indicators by themselves, which
are subject to large swings, but rather on temporal averages that have the
ability of smoothing out the noise inevitably present in the oil price
trajectory. Moreover, by the very design of the manufacturing production
chain, increases/decreases over several months are more likely to be
transmitted into prices than notoriously volatile one-month-to-the-next
variations.

\subsection{Quiet(er) Times}

It has been repeatedly reported that the benefits of a large panel of
predictors may solely be present during periods of economic turmoil %
\citep{kotchoni2019macroeconomic,bread}. For this reason and others %
\citep{Lerchetal(2017)}, it is of interest to study the marginal benefits
associated with data-rich models outside of the tumultuous entry/exit of the
Great Recession and the Pandemic Recession. Moreover, starting the
pseudo-out-of-sample from 2011 gives data-rich models \textit{at least one
recession to be trained on}, and 13 years of data overall rather than 10 (as
it were the case in Table \ref{tab:all}).

Ridge and Ridge-MARX do well for EMP and HOURS with gains roughly
distributed between 10\% and 20\% depending on the horizon. The MARX version
usually has the upper hand by a small margin. The evidence for other
activity indicators is more mixed. For HOURS, only nonlinear models manage
to beat the AR benchmark albeit in a non-statistically significant fashion.
The best model for IP PROD at all horizons is ARRF(2) which improves upon
the AR by small margins. For IP MACH, some small gains can be obtained at a
horizon of 3 months (with FA-ARRF(2,2), most notably) but none of those are
statistically significant.

Aligned with traditional wisdom for the US \citep{stock2008phillips}, it is
hard to beat the simple benchmark when it comes to CPI inflation.
Nevertheless, ARRF(6) is the best model for all horizons (ex-aequo at $h=1$)
with gains of 9-10\% -- but none of those are significant. Larger
improvements are obtained for RPI, where various data-rich models (linear
and nonlinear) provide gains of around 20\%. The most notable are those of
FA-ARRFs at a horizon of 3 months (but also any other horizon) which are
nearly 30\%, far ahead from most of the competing models -- including all
those that also rely directly on factors. Finally, as a last notable
observation from Table \ref{tab:quiet}, ARRF(6) dominates at all horizons
for both RPI HOUSING and CREDIT, highlighting the benefits of a more focused
modeling of persistence (while allowing for its time variation) in otherwise
high-dimensional/data-rich/nonlinear ML methods.




\section{Conclusion}

\label{sec:conclusion}

In this paper we assess the forecasting performance of a variety of standard
and ML forecasting methods for key UK economic variables, with a special
focus on the Covid-19 period and using a specifically collected large
dataset of monthly indicators, labeled UK-MD (also augmented with some
international indicators).

As standard benchmarks, we consider AR, random walk and factor augmented AR
models. As ML methods, we evaluate penalized regressions (RIDGE, LASSO,
ELASTIC NET), boosted trees (BT) and random forests (RF), Kernel Ridge
Regression (KRR), and Neural Networks (NN), plus Macroeconomic Random Forest
(MRF), which uses a linear part within the leafs, and Moving Average
Rotation of $X$ (MARX), a feature engineering technique which generates an
implicit shrinkage more appropriate for time series data.

Overall ML methods can provide substantial gains when short-term forecasting
several indicators of the UK economy, though a careful temporal and variable
by variable analysis is needed. Over the full sample, RF works particularly
well for labour market variables, in particular when augmented with MARX;
KRR for real activity and consumer price inflation; LASSO or LASSO+MARX for
the retail price index and its version focusing on housing; and RF for
credit variables. The gains can be sizable, even 40-50\% with respect to the
benchmark, and ML methods were particularly useful during the Covid-19
period. During the Covid era, nonlinear methods with the ability to
extrapolate have a nice edge. Certain MRFs, unlike linear methods or simpler
nonlinear ML techniques, procure important improvements by predicting large
"bounce back" that did occur and avoid predicting mayhem that did not
materialize.

\clearpage
\onehalfspace

\setlength\bibsep{5pt}
\bibliographystyle{apalike} 
\bibliography{references.bib}

\clearpage
\appendix


\section{Detailed Forecasting Results}

\label{sec:detailed}

\begin{landscape}
	\begin{table}[htp]
		\centering
		\footnotesize
		\caption{All Sample (2008-2020)}\label{tab:all}
		\setlength{\tabcolsep}{0.5em} %
		\setstretch{1.45}
		\rowcolors{9}{gray!15}{white}
		\begin{tabular}{l *{18}{l}}
			\toprule
			\hspace*{0.5cm} &
			\multicolumn{3}{c}{EMP} &
			\multicolumn{3}{c}{UNRATE} &
			\multicolumn{3}{c}{HOURS} & 
			\multicolumn{3}{c}{IP} &
			\multicolumn{3}{c}{IP MACH} &
			\multicolumn{3}{c}{RETAIL} \\
			
			\cmidrule(lr){2-4} \cmidrule(lr){5-7} \cmidrule(lr){8-10} \cmidrule(lr){11-13} \cmidrule(lr){14-16} \cmidrule(lr){17-19}
			& h=1 & h=2 & h=3 & h=1 & h=2 & h=3 & h=1 & h=2 & h=3 & h=1 & h=2 & h=3 & h=1 & h=2 & h=3 & h=1 & h=2 & h=3 \\ 
			\midrule
			RW & 1.30*** & 1.23* & 1.18 & 1.25*** & 1.24 & 1.18 & 1.46 & 0.83 & \textbf{0.91} & 0.76* & 0.93 & 1.00 & \textbf{0.82}** & 0.87 & 0.79 & 0.70* & 0.79 & 1.08 \\ 
			ARDI,BIC & 1.54*** & 1.13 & 1.03 & 1.34* & 1.37 & 1.07 & 1.85 & 0.85 & 0.93 & 1.63 & 1.07 & 1.09 & 1.64 & 0.92 & 0.84 & 1.22 & 0.79 & 1.07 \\ 
			\midrule
			LASSO & 1.30 & 1.45 & 1.97 & 1.43 & 1.42 & 1.98 & 1.60 & 0.89 & 0.94 & 1.73 & 0.99 & 1.02 & 1.71 & 0.93 & 0.85 & 0.73* & 0.82 & \textbf{1.01} \\ 
			LASSO+MARX & 1.28 & 1.62 & 2.18 & 1.42 & 1.56 & 1.78 & 1.67 & 0.92 & 0.98 & 1.97 & 0.93 & 1.04 & 1.74 & 1.04 & 0.85 & 0.76 & 0.85 & 1.09 \\ 
			RIDGE & 1.02 & 0.97 & 0.89 & 0.98 & 0.96 & 0.88 & 1.66 & \textbf{0.82} & 0.93 & \textbf{0.68}** & 1.08 & 1.13 & 1.05 & 0.99 & 1.11 & 0.72* & 0.91 & 1.26* \\ 
			RIDGE+MARX & 1.11 & 1.20 & 1.40 & 1.14 & 1.44 & 1.47 & 1.24 & 0.83 & 0.92 & 1.98 & 1.09 & 1.19 & 0.90 & 1.05 & 1.02 & 0.75 & 1.02 & 1.22 \\ 
			E-NET & 1.31 & 1.29 & 1.73 & 1.37 & 1.44 & 1.84 & 1.64 & 0.87 & 0.93 & 1.58 & 0.99 & 1.02 & 1.63 & 0.93 & 0.91 & 0.79 & 0.82 & 1.04 \\ 
			E-NET+MARX & 1.25 & 1.63 & 2.39 & 1.28 & 1.64 & 1.66 & 1.66 & 0.92 & 0.99 & 1.92 & 1.08 & 1.04 & 1.67 & 0.96 & 0.85 & 0.82 & 0.92 & 1.10** \\
			\midrule
			KRR-ARDI & 1.17** & 1.09 & 1.05 & 1.15* & 1.12 & 1.09 & 1.47 & 0.84 & 0.95 & 0.76* & 0.94 & 1.01 & 0.82** & \textbf{0.86} & \textbf{0.77} & \textbf{0.69}* & \textbf{0.79} & 1.06 \\ 
			RF & 1.01 & 0.92 & 0.86 & 0.88** & 0.82* & 0.82 & 1.33 & 1.00 & 1.03 & 0.94 & 1.29 & 1.18 & 1.03 & 1.19 & 0.92 & 0.86 & 0.97 & 1.11 \\ 
			RF+MARX & 0.96 & \textbf{0.85}** & \textbf{0.81}** & \textbf{0.83}*** & \textbf{0.73}** & \textbf{0.75}* & 1.22 & 1.04 & 1.07 & 1.00 & 1.62 & 1.18 & 1.11 & 1.42 & 0.92 & 0.95 & 1.22 & 1.15 \\ 
			Boosting & 1.05 & 0.97 & 0.92 & 1.00 & 0.96 & 0.95 & 1.41 & 0.84 & 0.94 & 0.76* & 0.95 & 1.04 & 0.83** & 0.90 & 0.81 & 0.71* & 0.80 & 1.08 \\ 
			Boosting+MARX & 1.04 & 0.92 & 0.87*** & 0.95 & 0.89 & 0.87 & 1.40 & 0.85 & 0.95 & 0.76* & 0.96 & 1.06 & 0.83** & 0.91 & 0.82 & 0.72* & 0.81 & 1.09 \\ 
			ARRF,2Ylag & \textbf{0.96} & 0.88** & 0.88* & 0.92* & 0.82* & 0.83 & 0.79 & 1.12 & 1.41 & 1.52 & \textbf{0.85} & 1.22 & 1.92 & 1.76 & 1.22 & 1.70 & 0.91 & 2.09 \\ 
			FA-ARRF,2Fac & 0.98 & 1.09 & 1.19 & 1.14 & 1.60 & 1.68 & 0.72 & 0.98 & 4.61 & 1.38 & 1.05 & 0.93 & 2.46 & 2.71 & 1.49 & 2.13 & 1.16 & 1.37 \\ 
			ARRF,6Ylag & 1.01 & 0.94 & 0.98 & 0.93 & 0.88 & 0.93 & 0.79 & 0.93 & 1.49 & 1.23 & 0.92 & 1.19 & 1.73 & 2.93 & 1.95 & 1.02 & 0.99 & 4.48 \\ 
			FA-ARRF,4Fac & 0.99 & 1.00 & 1.04 & 1.01 & 1.37 & 1.02 & \textbf{0.70} & 1.02 & 2.54 & 1.41 & 1.09 & \textbf{0.82} & 2.73 & 1.34 & 1.16 & 1.84 & 1.08 & 1.32 \\ 
			NN-ARDI & 1.07 & 0.97 & 0.90* & 1.05 & 0.87 & 0.84 & 1.04 & 0.92 & 1.05 & 0.75** & 0.93 & 0.98 & 1.01 & 0.88 & 0.81 & 0.74 & 0.79 & 1.05 \\ 
			NN-ARDI+MARX & 1.32** & 1.14 & 0.94 & 1.16 & 1.08 & 0.92 & 1.55 & 1.25 & 1.40 & 1.44 & 2.17 & 1.91 & 2.36 & 2.23 & 1.02 & 1.30 & 1.03 & 1.49 \\ 
			\toprule
		\end{tabular}
		\begin{tablenotes}[para,flushleft]
			Notes: The numbers represent the relative MSEs with respect to AR,BIC model. 
$^{***}$. $^{**}$. $^{*}$ stand for 1\%. 5\% and 10\% significance of Diebold-Mariano test.
		\end{tablenotes}
	\end{table}
\end{landscape}

\begin{landscape}
	\begin{table}[htp]
		\centering
		\footnotesize
		\caption{All Sample (2008-2020), Continued}
		\setlength{\tabcolsep}{0.3454em} %
		\setstretch{1.45}
		\rowcolors{9}{gray!15}{white}
		
		\begin{tabular}{l *{18}{l}}
			\toprule
			\hspace*{0.5cm} &
			\multicolumn{3}{c}{CPI} &
			\multicolumn{3}{c}{RPI} &
			\multicolumn{3}{c}{RPI HOUSING} &
			\multicolumn{3}{c}{CREDIT} &
			\multicolumn{3}{c}{HOUSE APP} &
			\multicolumn{3}{c}{PPI MANU} \\
			
			\cmidrule(lr){2-4} \cmidrule(lr){5-7} \cmidrule(lr){8-10} \cmidrule(lr){11-13} \cmidrule(lr){14-16} \cmidrule(lr){17-19}
			& h=1 & h=2 & h=3 & h=1 & h=2 & h=3 & h=1 & h=2 & h=3 & h=1 & h=2 & h=3 & h=1 & h=2 & h=3 & h=1 & h=2 & h=3 \\ 
			\midrule
			RW & 2.88*** & 4.03*** & 4.97*** & 1.71*** & 2.02*** & 2.27*** & 1.37* & 1.39 & 1.30 & 0.90 & 0.88 & 0.91 & 0.68 & 0.62 & 0.84 & 1.71*** & 1.40** & 1.26 \\ 
			ARDI,BIC & 1.35 & 1.62* & 1.96*** & 1.84*** & 2.22** & 1.74** & 2.75** & 3.23* & 3.10* & 1.45 & 1.10 & 1.34 & 1.15 & 0.69 & 0.91 & 2.66*** & 2.17** & 1.56** \\ 
			\midrule
			LASSO & 1.08 & 1.11 & 1.33 & 0.76 & 0.95 & 1.10 & 0.43 & \textbf{0.82} & 1.25 & 1.03 & 0.84 & 0.91 & 0.82 & 0.58 & 0.87 & 0.96 & 1.13 & 1.16 \\ 
			LASSO+MARX & 1.10 & 1.09 & 1.24* & 0.77* & 1.15 & 1.14 & \textbf{0.41} & 0.85 & 1.12 & 1.12 & 0.88 & 0.91 & 0.70 & 0.70 & 0.86 & 1.01 & 1.18 & 1.25 \\ 
			RIDGE & 1.35 & 1.23 & 1.33** & 1.16 & 1.29 & 1.40 & 1.10 & 1.74 & 1.61 & 0.93 & 0.91 & 0.96 & 0.81 & \textbf{0.56} & 0.92 & 1.35*** & 1.44** & 1.39** \\ 
			RIDGE+MARX & 1.23 & 1.22 & 1.31 & 0.97 & 1.22 & 1.55 & 0.79 & 1.79 & 2.01 & 1.03 & 1.01 & 1.12 & 0.77 & 0.64 & 0.96 & 1.25 & 1.30 & 1.45* \\ 
			E-NET & 1.26 & 1.22 & 1.20 & 0.88 & \textbf{0.93} & 1.09 & 0.48 & 0.91 & 1.21 & 1.00 & 0.85 & 0.91 & 0.75 & 0.60 & 0.87 & \textbf{0.95} & 1.19 & 1.26 \\ 
			E-NET+MARX & 1.10 & 1.10 & 1.21 & 0.88 & 0.98 & 1.12 & 0.49 & 0.98 & 1.14 & 1.08 & 0.87 & 0.87 & 0.86 & 0.75 & 0.90** & 1.00 & 1.12 & 1.25 \\ 
			\midrule
			KRR-ARDI & \textbf{0.87} & 0.86 & \textbf{0.96} & 0.99 & 1.02 & 1.06 & 1.50** & 1.65 & 1.62 & 0.95 & 0.93 & 0.98 & 0.69 & 0.63 & 0.89 & 1.28** & 1.15 & \textbf{1.11} \\ 
			RF & 0.93 & 0.89 & 1.05 & 0.89** & 1.01 & 1.13 & 0.87 & 1.12 & 1.21* & 0.81** & \textbf{0.76}* & \textbf{0.84} & \textbf{0.66}* & 0.82 & 1.05* & 1.18* & 1.32 & 1.42 \\ 
			RF+MARX & 0.91 & \textbf{0.85} & 1.00 & 0.87** & 1.01 & 1.16 & 0.80 & 1.10 & 1.20* & \textbf{0.80}** & 0.79 & 0.84 & 0.70 & 1.01 & 1.08 & 1.10 & 1.25 & 1.39 \\ 
			Boosting & 0.97 & 1.02 & 1.15 & 0.98 & 1.00 & \textbf{1.04} & 1.15 & 1.28 & 1.28 & 0.84* & 0.82 & 0.88 & 0.67 & 0.63 & 0.87 & 1.27** & 1.22 & 1.16 \\ 
			Boosting+MARX & 0.96 & 1.01 & 1.12 & 0.97 & 1.00 & 1.05 & 1.14 & 1.27 & 1.27 & 0.84* & 0.82 & 0.89 & 0.68 & 0.63 & 0.87 & 1.24** & 1.21 & 1.17 \\ 
			ARRF,2Ylag & 1.51 & 1.13 & 0.97 & 0.99 & 1.01 & 1.17 & 1.20 & 1.00 & \textbf{1.08} & 0.95 & 0.95 & 1.00 & 0.74 & 1.39 & 5.48 & 0.99 & \textbf{1.07} & 1.33 \\ 
			FA-ARRF,2Fac & 1.32 & 1.33* & 1.69 & 0.82 & 1.06 & 1.67 & 1.40 & 4.39 & 4.11 & 1.00 & 1.02 & 1.16 & 0.92 & 0.72 & 2.71 & 1.13 & 2.09 & 2.47 \\ 
			ARRF,6Ylag & 1.33 & 1.18 & 1.15 & 1.01 & 1.19 & 1.30 & 1.21 & 0.97 & 1.58 & 1.13 & 1.08 & 1.24* & 0.92 & 0.72 & 3.62 & 1.11 & 1.20 & 1.59 \\ 
			FA-ARRF,4Fac & 1.37 & 1.41* & 1.78 & \textbf{0.74} & 1.31 & 1.64 & 1.11 & 1.89 & 2.45 & 1.06 & 0.99 & 0.99 & 0.69 & 0.64 & \textbf{0.79} & 1.11 & 2.42 & 3.52 \\ 
			NN-ARDI & 1.03 & 0.90 & 1.19 & 1.04 & 1.06 & 1.20 & 0.92 & 1.46 & 1.53 & 0.89 & 0.93 & 0.92 & 0.66* & 0.62 & 0.84 & 1.36*** & 1.18 & 1.22 \\ 
			NN-ARDI+MARX & 1.12 & 1.10 & 1.09 & 0.90 & 1.21 & 1.68 & 0.99 & 1.51 & 1.88 & 1.27 & 1.22 & 1.27 & 0.74 & 1.03 & 0.86 & 1.39** & 1.24 & 2.16 \\ 
			\toprule
		\end{tabular}
		\begin{tablenotes}[para,flushleft]
			Notes: See Table \ref{tab:all}.
		\end{tablenotes}
	\end{table}
\end{landscape}

\begin{landscape}
	\begin{table}[htp]
		\begin{threeparttable}
			\begin{scriptsize}
				\centering
				\footnotesize
				\caption{Restricted Sample (2011-2020)}
				\setlength{\tabcolsep}{0.5em} %
				\setstretch{1.45}
				\rowcolors{9}{gray!15}{white}
				\begin{tabular}{l *{18}{l}}
					\toprule
					\hspace*{0.5cm} &
					\multicolumn{3}{c}{EMP} &
					\multicolumn{3}{c}{UNRATE} &
					\multicolumn{3}{c}{HOURS} & 
					\multicolumn{3}{c}{IP} &
					\multicolumn{3}{c}{IP MACH} &
					\multicolumn{3}{c}{RETAIL} \\
					
					\cmidrule(lr){2-4} \cmidrule(lr){5-7} \cmidrule(lr){8-10} \cmidrule(lr){11-13} \cmidrule(lr){14-16} \cmidrule(lr){17-19}
					& h=1 & h=2 & h=3 & h=1 & h=2 & h=3 & h=1 & h=2 & h=3 & h=1 & h=2 & h=3 & h=1 & h=2 & h=3 & h=1 & h=2 & h=3 \\ 
					\midrule
					RW &  1.40*** &  1.40* &  1.50 &  1.20*** &  1.32 &  1.34 &  1.49 &  0.83 & \textbf{0.91} &  0.75* &  0.93 &  1.02 &  0.83** & \textbf{0.87} &  0.77 &  0.67* &  0.78 &  1.10 \\ 
					ARDI,BIC &  1.51*** &  1.10 &  1.00 &  1.33* &  1.62 &  1.04 &  1.90 &  0.85 &  0.93 &  1.66 &  1.07 &  1.11 &  1.71 &  0.92 &  0.83 &  1.22 &  0.77 &  1.07 \\
					\midrule
					LASSO &  1.46 &  1.68 &  2.59 &  1.78 &  2.05 &  3.36 &  1.64 &  0.89 &  0.95 &  1.77 &  0.99 &  1.04 &  1.80 &  0.94 &  0.86 &  0.70* &  0.81 & \textbf{1.01} \\ 
					LASSO+MARX &  1.44 &  2.00 &  3.04 &  1.78 &  2.31 &  3.08 &  1.71 &  0.92 &  0.98 &  2.03 &  0.93 &  1.06 &  1.83 &  1.06 &  0.87 &  0.74 &  0.84 &  1.10 \\ 
					RIDGE &  1.08 &  1.06 &  1.01 &  1.10 &  1.32 &  1.32 &  1.71 & \textbf{0.82} &  0.93 & \textbf{0.67}** &  1.10 &  1.16 &  1.09 &  1.02 &  1.16 &  0.69* &  0.90 &  1.27* \\ 
					RIDGE+MARX &  1.21 &  1.44 &  1.90 &  1.38 &  2.28 &  2.59 &  1.27 &  0.84 &  0.93 &  2.05 &  1.11 &  1.24 &  0.92 &  1.09 &  1.06 &  0.72 &  1.02 &  1.24 \\ 
					ENET &  1.45 &  1.42 &  2.24 &  1.74 &  2.07 &  3.15 &  1.69 &  0.87 &  0.93 &  1.61 &  1.00 &  1.04 &  1.71 &  0.94 &  0.93 &  0.77 &  0.80 &  1.05 \\ 
					E-NET+MARX &  1.41 &  2.03 &  3.37 &  1.57 &  2.51 &  2.88 &  1.71 &  0.92 &  1.00 &  1.98 &  1.10 &  1.06 &  1.75 &  0.98 &  0.86 &  0.81 &  0.91 &  1.10** \\ 
					\midrule
					KRR-ARDI &  1.12** &  1.05 &  1.09 &  1.07* &  1.12 &  1.15 &  1.50 &  0.84 &  0.95 &  0.75* &  0.94 &  1.04 & \textbf{0.82}** &  0.87 & \textbf{0.76} & \textbf{0.66}* & \textbf{0.77} &  1.07 \\ 
					RF &  1.03 &  0.95 &  0.92 &  0.97** &  0.97* &  1.00 &  1.36 &  1.00 &  1.03 &  0.94 &  1.32 &  1.21 &  1.06 &  1.23 &  0.93 &  0.84 &  0.96 &  1.12 \\ 
					RF+MARX &  0.99 & \textbf{0.88}** & \textbf{0.82}** & \textbf{0.94}*** & \textbf{0.88}** & \textbf{0.84}* &  1.24 &  1.04 &  1.08 &  1.01 &  1.68 &  1.21 &  1.14 &  1.48 &  0.93 &  0.94 &  1.22 &  1.16 \\ 
					Boosting &  1.03 &  0.94 &  0.93 &  1.02 &  1.05 &  1.03 &  1.44 &  0.84 &  0.94 &  0.76* &  0.95 &  1.06 &  0.84** &  0.90 &  0.80 &  0.69* &  0.79 &  1.08 \\ 
					Boosting+MARX &  1.02 &  0.89 &  0.86*** &  0.99 &  0.99 &  0.96 &  1.44 &  0.85 &  0.95 &  0.76* &  0.97 &  1.08 &  0.84** &  0.91 &  0.82 &  0.69* &  0.80 &  1.09 \\ 
					ARRF,2Ylag &  0.97 &  0.88** &  0.85* &  0.98* &  0.93* &  0.94 &  0.78 &  1.13 &  1.42 &  1.55 & \textbf{0.84} &  1.23 &  2.00 &  1.83 &  1.25 &  1.75 &  0.91 &  2.16 \\ 
					FA-ARRF,2Fac &  0.96 &  1.13 &  1.26 &  1.28 &  2.34 &  2.66 &  0.72 &  0.98 &  4.70 &  1.41 &  1.07 &  0.93 &  2.61 &  2.90 &  1.57 &  2.21 &  1.16 &  1.40 \\ 
					ARRF,6Ylag &  1.02 &  0.95 &  0.92 &  0.97 &  0.99 &  1.02 &  0.78 &  0.94 &  1.50 &  1.24 &  0.90 &  1.18 &  1.80 &  3.12 &  2.07 &  1.03 &  0.99 &  4.69 \\ 
					FA-ARRF,4Fac & \textbf{0.95} &  0.97 &  1.01 &  1.08 &  1.89 &  1.33 & \textbf{0.69} &  1.03 &  2.58 &  1.43 &  1.11 & \textbf{0.80} &  2.90 &  1.40 &  1.21 &  1.90 &  1.08 &  1.35 \\ 
					NN-ARDI &  1.04 &  1.00 &  0.95* &  1.04 &  1.02 &  1.03 &  1.05 &  0.92 &  1.05 &  0.74** &  0.93 &  0.99 &  1.03 &  0.89 &  0.81 &  0.72 &  0.77 &  1.05 \\ 
					NN-ARDI+MARX &  1.46** &  1.27 &  1.12 &  1.30 &  1.65 &  1.29 &  1.58 &  1.25 &  1.42 &  1.47 &  2.26 &  2.05 &  2.51 &  2.38 &  1.05 &  1.30 &  1.03 &  1.52 \\ 
					\toprule				
					
				\end{tabular}
				\begin{tablenotes}[para,flushleft]
					Notes: See Table \ref{tab:all}.
				\end{tablenotes}
			\end{scriptsize}
		\end{threeparttable}
	\end{table}
\end{landscape}

\begin{landscape}
	\begin{scriptsize}
		\begin{table}[htp]
			\begin{threeparttable}
				\centering
				\footnotesize
				\caption{Restricted Sample (2011-2020), Continued}
				\setlength{\tabcolsep}{0.32em} %
				\setstretch{1.45}
				\rowcolors{9}{gray!15}{white}
				\begin{tabular}{l *{18}{l}}
					\toprule
					\hspace*{0.5cm} &
					\multicolumn{3}{c}{CPI} &
					\multicolumn{3}{c}{RPI} &
					\multicolumn{3}{c}{RPI HOUSING} &
					\multicolumn{3}{c}{CREDIT} &
					\multicolumn{3}{c}{HOUSE APP} &
					\multicolumn{3}{c}{PPI MANU} \\
					
					\cmidrule(lr){2-4} \cmidrule(lr){5-7} \cmidrule(lr){8-10} \cmidrule(lr){11-13} \cmidrule(lr){14-16} \cmidrule(lr){17-19}
					& h=1 & h=2 & h=3 & h=1 & h=2 & h=3 & h=1 & h=2 & h=3 & h=1 & h=2 & h=3 & h=1 & h=2 & h=3 & h=1 & h=2 & h=3 \\ 
					\midrule
					RW &  3.03*** &  4.98*** &  6.23*** &  1.87*** &  2.96*** &  3.64*** &  1.85* &  3.28 &  3.89 &  1.46 &  1.51 &  1.69 &  0.66 &  0.59 &  0.82 &  1.42*** &  1.19** &  1.17 \\ 
					ARDI,BIC &  0.97 &  1.40* &  2.05*** &  1.28*** &  2.34** &  1.77** &  3.29** & 11.50* & 10.22* &  3.20 &  1.53 &  2.35 &  1.14 &  0.64 &  0.83 &  1.79*** &  1.97** &  1.34** \\ 
					\midrule
					LASSO &  1.14 &  1.25 &  1.50 &  0.90 &  0.76 &  0.88 &  2.39 &  5.48 & 11.22 &  1.73 &  1.26 &  1.59 &  0.80 &  0.56 &  0.86 & \textbf{0.91} &  1.05 &  1.09 \\ 
					LASSO+MARX &  1.18 &  1.22 &  1.32* &  0.82* &  0.87 &  1.03 &  2.28 &  5.70 &  8.81 &  1.83 &  1.38 &  1.55 &  0.68 &  0.69 &  0.86 &  0.94 &  1.05 &  1.12 \\ 
					RIDGE &  1.70 &  1.48 &  1.37** &  1.36 &  1.51 &  1.49 &  4.45 & 16.54 & 16.48 &  1.44 &  1.53 &  1.71 &  0.80 & \textbf{0.55} &  0.93 &  1.37*** &  1.48** &  1.27** \\ 
					RIDGE+MARX &  1.47 &  1.49 &  1.49 &  1.16 &  1.35 &  2.15 &  3.08 & 18.27 & 26.05 &  1.60 &  1.62 &  2.00 &  0.77 &  0.64 &  0.98 &  1.31 &  1.40 &  1.52* \\ 
					ENET &  1.53 &  1.47 &  1.21 &  0.90 & \textbf{0.74} &  0.88 &  2.53 &  6.80 & 10.18 &  1.50 &  1.32 &  1.56 &  0.73 &  0.59 &  0.87 &  0.94 &  1.06 &  1.09 \\ 
					E-NET+MARX &  1.16 &  1.17 &  1.38 &  0.97 &  0.82 &  0.97 &  2.48 &  7.44 &  9.06 &  1.68 &  1.30 &  1.37 &  0.84 &  0.74 &  0.90** &  0.94 &  1.05 &  1.09 \\ 
					\midrule
					KRR-ARDI & \textbf{0.87} &  0.88 &  0.93 &  0.82 &  0.86 &  0.85 &  1.51** &  3.19 &  4.10 &  1.25 &  1.14 &  1.19 &  0.67 &  0.61 &  0.86 &  0.96** & \textbf{0.85} & \textbf{0.84} \\ 
					RF &  1.02 &  0.95 &  1.03 &  0.79** &  0.79 &  0.79 &  1.18 &  1.73 &  1.89* &  0.91** &  0.97* &  1.07 & \textbf{0.64}* &  0.81 &  1.01* &  0.97* &  0.99 &  1.02 \\ 
					RF+MARX &  1.00 &  0.92 &  1.00 &  0.78** &  0.75 &  0.77 &  1.57 &  2.41 &  2.34* & \textbf{0.81}** &  0.97 &  1.08 &  0.68 &  1.00 &  1.05 &  0.99 &  1.01 &  1.07 \\ 
					Boosting &  1.01 &  1.15 &  1.28 &  0.79 &  0.77 &  0.77 &  0.85 &  0.75 &  0.86 &  1.05* &  1.06 &  1.17 &  0.65 &  0.61 &  0.84 &  1.03** &  1.02 &  1.00 \\ 
					Boosting+MARX &  0.99 &  1.14 &  1.28 & \textbf{0.77} &  0.74 & \textbf{0.76} &  0.85 &  0.74 &  0.83 &  1.02* &  1.03 &  1.15 &  0.66 &  0.61 &  0.85 &  1.00** &  1.00 &  0.99 \\ 
					ARRF,2Ylag &  0.96 &  1.05 &  0.99 &  0.85 &  0.81 &  0.83 &  0.40 &  1.26 & \textbf{0.72} &  0.93 &  0.92 & \textbf{0.94} &  0.73 &  1.40 &  5.87 &  0.94 &  0.98 &  1.03 \\ 
					FA-ARRF,2Fac &  0.97 &  1.42* &  1.95 &  0.89 &  1.17 &  2.52 &  5.91 & 59.67 & 65.78 &  1.15 &  1.27 &  1.58 &  0.90 &  0.71 &  2.84 &  1.03 &  2.82 &  3.12 \\ 
					ARRF,6Ylag &  0.92 &  0.96 & \textbf{0.89} &  0.87 &  0.87 &  0.84 & \textbf{0.30} & \textbf{0.53} &  0.99 &  0.90 & \textbf{0.88} &  0.98* &  0.92 &  0.69 &  3.79 &  0.94 &  0.93 &  0.96 \\ 
					FA-ARRF,4Fac &  1.00 &  1.41* &  1.89 &  0.82 &  1.77 &  2.38 &  2.90 & 16.29 & 29.22 &  1.25 &  1.14 &  1.07 &  0.67 &  0.63 & \textbf{0.77} &  1.03 &  3.39 &  5.59 \\ 
					NN-ARDI &  1.11 & \textbf{0.87} &  1.26 &  0.92 &  1.05 &  1.21 &  1.31 &  8.10 & 11.57 &  1.03 &  1.16 &  1.28 &  0.64* &  0.60 &  0.82 &  1.23*** &  1.03 &  1.11 \\ 
					NN-ARDI+MARX &  1.14 &  1.37 &  1.12 &  1.07 &  1.23 &  2.42 &  4.72 & 16.68 & 21.54 &  2.35 &  1.87 &  2.02 &  0.73 &  1.04 &  0.88 &  1.35** &  1.32 &  2.99 \\ 
					\toprule				
				\end{tabular}
				\begin{tablenotes}[para,flushleft]
					Notes: See Table \ref{tab:all}.
				\end{tablenotes}
			\end{threeparttable}
		\end{table}
	\end{scriptsize}
\end{landscape}

\begin{landscape}
	\begin{scriptsize}
		\begin{table}[htp]
			\begin{threeparttable}
				\centering
				\footnotesize
				\caption{Covid Sample (from 2020m1)}
				\setlength{\tabcolsep}{0.50em} %
				\setstretch{1.45}
				\rowcolors{9}{gray!15}{white}
				\begin{tabular}{l *{18}{l}}
					\toprule
					\hspace*{0.5cm} &
					\multicolumn{3}{c}{EMP} &
					\multicolumn{3}{c}{UNRATE} &
					\multicolumn{3}{c}{HOURS} & 
					\multicolumn{3}{c}{IP} &
					\multicolumn{3}{c}{IP MACH} &
					\multicolumn{3}{c}{RETAIL} \\
					
					\cmidrule(lr){2-4} \cmidrule(lr){5-7} \cmidrule(lr){8-10} \cmidrule(lr){11-13} \cmidrule(lr){14-16} \cmidrule(lr){17-19}
					& h=1 & h=2 & h=3 & h=1 & h=2 & h=3 & h=1 & h=2 & h=3 & h=1 & h=2 & h=3 & h=1 & h=2 & h=3 & h=1 & h=2 & h=3 \\ 
					\midrule
					RW &  1.40*** &  1.40* &  1.50 &  1.20*** &  1.32 &  1.34 &  1.49 &  0.83 & \textbf{0.91} &  0.75* &  0.93 &  1.02 &  0.83** & \textbf{0.87} &  0.77 &  0.67* &  0.78 &  1.10 \\ 
					ARDI,BIC &  1.51*** &  1.10 &  1.00 &  1.33* &  1.62 &  1.04 &  1.90 &  0.85 &  0.93 &  1.66 &  1.07 &  1.11 &  1.71 &  0.92 &  0.83 &  1.22 &  0.77 &  1.07 \\
					\midrule
					LASSO &  1.46 &  1.68 &  2.59 &  1.78 &  2.05 &  3.36 &  1.64 &  0.89 &  0.95 &  1.77 &  0.99 &  1.04 &  1.80 &  0.94 &  0.86 &  0.70* &  0.81 & \textbf{1.01} \\ 
					LASSO+MARX &  1.44 &  2.00 &  3.04 &  1.78 &  2.31 &  3.08 &  1.71 &  0.92 &  0.98 &  2.03 &  0.93 &  1.06 &  1.83 &  1.06 &  0.87 &  0.74 &  0.84 &  1.10 \\ 
					RIDGE &  1.08 &  1.06 &  1.01 &  1.10 &  1.32 &  1.32 &  1.71 & \textbf{0.82} &  0.93 & \textbf{0.67}** &  1.10 &  1.16 &  1.09 &  1.02 &  1.16 &  0.69* &  0.90 &  1.27* \\ 
					RIDGE+MARX &  1.21 &  1.44 &  1.90 &  1.38 &  2.28 &  2.59 &  1.27 &  0.84 &  0.93 &  2.05 &  1.11 &  1.24 &  0.92 &  1.09 &  1.06 &  0.72 &  1.02 &  1.24 \\ 
					ENET &  1.45 &  1.42 &  2.24 &  1.74 &  2.07 &  3.15 &  1.69 &  0.87 &  0.93 &  1.61 &  1.00 &  1.04 &  1.71 &  0.94 &  0.93 &  0.77 &  0.80 &  1.05 \\ 
					E-NET+MARX &  1.41 &  2.03 &  3.37 &  1.57 &  2.51 &  2.88 &  1.71 &  0.92 &  1.00 &  1.98 &  1.10 &  1.06 &  1.75 &  0.98 &  0.86 &  0.81 &  0.91 &  1.10** \\ 
					\midrule
					KRR-ARDI &  1.12** &  1.05 &  1.09 &  1.07* &  1.12 &  1.15 &  1.50 &  0.84 &  0.95 &  0.75* &  0.94 &  1.04 & \textbf{0.82}** &  0.87 & \textbf{0.76} & \textbf{0.66}* & \textbf{0.77} &  1.07 \\ 
					RF &  1.03 &  0.95 &  0.92 &  0.97** &  0.97* &  1.00 &  1.36 &  1.00 &  1.03 &  0.94 &  1.32 &  1.21 &  1.06 &  1.23 &  0.93 &  0.84 &  0.96 &  1.12 \\ 
					RF+MARX &  0.99 & \textbf{0.88}** & \textbf{0.82}** & \textbf{0.94}*** & \textbf{0.88}** & \textbf{0.84}* &  1.24 &  1.04 &  1.08 &  1.01 &  1.68 &  1.21 &  1.14 &  1.48 &  0.93 &  0.94 &  1.22 &  1.16 \\ 
					Boosting &  1.03 &  0.94 &  0.93 &  1.02 &  1.05 &  1.03 &  1.44 &  0.84 &  0.94 &  0.76* &  0.95 &  1.06 &  0.84** &  0.90 &  0.80 &  0.69* &  0.79 &  1.08 \\ 
					Boosting+MARX &  1.02 &  0.89 &  0.86*** &  0.99 &  0.99 &  0.96 &  1.44 &  0.85 &  0.95 &  0.76* &  0.97 &  1.08 &  0.84** &  0.91 &  0.82 &  0.69* &  0.80 &  1.09 \\ 
					ARRF,2Ylag &  0.97 &  0.88** &  0.85* &  0.98* &  0.93* &  0.94 &  0.78 &  1.13 &  1.42 &  1.55 & \textbf{0.84} &  1.23 &  2.00 &  1.83 &  1.25 &  1.75 &  0.91 &  2.16 \\ 
					FA-ARRF,2Fac &  0.96 &  1.13 &  1.26 &  1.28 &  2.34 &  2.66 &  0.72 &  0.98 &  4.70 &  1.41 &  1.07 &  0.93 &  2.61 &  2.90 &  1.57 &  2.21 &  1.16 &  1.40 \\ 
					ARRF,6Ylag &  1.02 &  0.95 &  0.92 &  0.97 &  0.99 &  1.02 &  0.78 &  0.94 &  1.50 &  1.24 &  0.90 &  1.18 &  1.80 &  3.12 &  2.07 &  1.03 &  0.99 &  4.69 \\ 
					FA-ARRF,4Fac & \textbf{0.95} &  0.97 &  1.01 &  1.08 &  1.89 &  1.33 & \textbf{0.69} &  1.03 &  2.58 &  1.43 &  1.11 & \textbf{0.80} &  2.90 &  1.40 &  1.21 &  1.90 &  1.08 &  1.35 \\ 
					NN-ARDI &  1.04 &  1.00 &  0.95* &  1.04 &  1.02 &  1.03 &  1.05 &  0.92 &  1.05 &  0.74** &  0.93 &  0.99 &  1.03 &  0.89 &  0.81 &  0.72 &  0.77 &  1.05 \\ 
					NN-ARDI+MARX &  1.46** &  1.27 &  1.12 &  1.30 &  1.65 &  1.29 &  1.58 &  1.25 &  1.42 &  1.47 &  2.26 &  2.05 &  2.51 &  2.38 &  1.05 &  1.30 &  1.03 &  1.52 \\ 
					\toprule				
					
				\end{tabular}
				\begin{tablenotes}[para,flushleft]
					Notes: See Table \ref{tab:all}.
				\end{tablenotes}
			\end{threeparttable}
		\end{table}
	\end{scriptsize}
\end{landscape}

\begin{landscape}
	\begin{table}[htp]
		\begin{threeparttable}
			\begin{scriptsize}
				\centering
				\footnotesize
				\caption{Covid Sample (from 2020m1), Continued}
				\setlength{\tabcolsep}{0.295em} %
				\setstretch{1.45}
				\rowcolors{9}{gray!15}{white}
				\begin{tabular}{ l *{18}{l}}
					\toprule
					\hspace*{0.5cm} &
					\multicolumn{3}{c}{CPI} &
					\multicolumn{3}{c}{RPI} &
					\multicolumn{3}{c}{RPI HOUSING} &
					\multicolumn{3}{c}{CREDIT} &
					\multicolumn{3}{c}{HOUSE APP} &
					\multicolumn{3}{c}{PPI MANU} \\
					
					\cmidrule(lr){2-4} \cmidrule(lr){5-7} \cmidrule(lr){8-10} \cmidrule(lr){11-13} \cmidrule(lr){14-16} \cmidrule(lr){17-19}
					& h=1 & h=2 & h=3 & h=1 & h=2 & h=3 & h=1 & h=2 & h=3 & h=1 & h=2 & h=3 & h=1 & h=2 & h=3 & h=1 & h=2 & h=3 \\ 
					\midrule
					RW & \textbf{0.68}*** & \textbf{0.54}*** & \textbf{0.45}*** &   0.67*** &   0.64*** &   0.58*** &   2.16* &   2.01 &   2.74 &   0.99 & \textbf{0.88} &   0.89 &   0.66 &   0.59 &   0.82 &   1.18*** &   0.81** &   0.66 \\ 
					ARDI,BIC &   0.70 &   1.09* &   2.39*** &   1.42*** &   6.19** &   3.09** &  19.44** &  66.21* &  56.23* &   4.23 &   1.53 &   2.65 &   1.14 &   0.63 &   0.81 &   3.78*** &   6.07** &   1.67** \\
					\midrule
					LASSO &   1.08 &   1.28 &   1.78 &   1.19 &   0.65 &   1.31 &  19.63 &  32.53 & 116.35 &   1.72 &   1.15 &   1.58 &   0.80 &   0.56 &   0.86 &   1.00 &   1.29 &   1.12 \\ 
					LASSO+MARX &   1.12 &   1.26 &   1.63* &   0.95* &   1.20 &   2.12 &  17.37 &  35.02 &  88.65 &   1.76 &   1.32 &   1.45 &   0.67 &   0.68 &   0.85 &   1.02 &   1.47 &   1.23 \\ 
					RIDGE &   1.89 &   1.63 &   1.45** &   2.86 &   4.48 &   4.52 &  58.87 & 140.96 & 186.01 &   1.41 &   1.55 &   1.74 &   0.80 & \textbf{0.54} &   0.93 &   3.00*** &   4.03** &   2.48** \\ 
					RIDGE+MARX &   1.62 &   1.76 &   1.94 &   2.22 &   3.94 &   8.55 &  35.49 & 157.60 & 316.57 &   1.59 &   1.63 &   2.17 &   0.76 &   0.63 &   0.97 &   3.11 &   3.56 &   4.41* \\ 
					E-NET &   1.63 &   1.71 &   1.36 &   1.17 & \textbf{0.55} &   1.17 &  24.38 &  47.22 & 101.36 &   1.32 &   1.24 &   1.51 &   0.73 &   0.58 &   0.87 &   1.03 &   1.52 &   1.08 \\ 
					E-NET+MARX &   1.13 &   1.16 &   1.70 &   1.44 &   0.97 &   1.78 &  24.15 &  51.50 &  88.89 &   1.69 &   1.22 &   1.24 &   0.84 &   0.74 &   0.89** &   0.88 &   1.35 &   1.02 \\ 
					\midrule
					KRR-ARDI &   0.83 &   0.83 &   0.92 &   0.76 &   0.81 &   0.60 &   1.62** &   1.42 &   1.75 &   1.21 &   1.08 &   1.17 &   0.66 &   0.60 &   0.85 &   1.10** & \textbf{0.57} & \textbf{0.35} \\ 
					RF &   0.96 &   0.73 &   0.70 &   0.75** &   0.82 &   0.60 &   9.84 &  12.29 &  17.95* &   0.90** &   0.99* &   1.10 & \textbf{0.63}* &   0.80 &   1.01* &   0.91* &   0.92 &   0.99 \\ 
					RF+MARX &   0.97 &   0.72 &   0.71 &   0.73** &   0.61 & \textbf{0.48} &  15.73 &  18.62 &  22.43* & \textbf{  0.72}** &   0.98 &   1.12 &   0.67 &   0.99 &   1.04 & \textbf{0.55} &   0.72 &   0.83 \\ 
					Boosting &   0.81 &   0.77 &   0.82 &   0.67 &   0.56 &   0.51 &   1.40 & \textbf{1.05} & \textbf{1.50} &   1.07* &   1.06 &   1.17 &   0.64 &   0.60 &   0.84 &   0.95** &   0.81 &   0.75 \\ 
					Boosting+MARX &   0.81 &   0.78 &   0.82 & \textbf{0.66} &   0.55 &   0.53 &   1.54 &   1.12 &   1.61 &   1.02* &   1.02 &   1.15 &   0.65 &   0.61 &   0.84 &   0.83** &   0.74 &  0.73 \\ 
					ARRF,2Ylag &   0.97 &   1.10 &   1.00 &   0.78 &   0.67 &   0.65 &   0.75 &   7.97 &   2.14 &   0.84 &   0.88 & \textbf{0.89} &   0.72 &   1.41 &   6.01 &   0.95 &   0.98 &   0.92 \\ 
					FA-ARRF,2Fac &   0.99 &   1.70* &   3.20 &   1.16 &   3.08 &  11.02 & 105.58 & 594.33 & 898.11 &   1.13 &   1.29 &   1.70 &   0.90 &   0.70 &   2.89 &   1.80 &  13.89 &  16.43 \\ 
					ARRF,6Ylag &   0.92 &   1.00 &   0.88 &   0.82 &   0.82 &   0.76 & \textbf{0.54} &   1.19 &   7.88 &   0.89 &   0.91 &   1.08* &   0.91 &   0.69 &   3.87 &   0.97 &   0.89 &   0.96 \\ 
					FA-ARRF,4Fac &   1.03 &   1.77* &   3.19 &   0.96 &   6.27 &  10.29 &  44.57 & 151.62 & 386.12 &   1.26 &   1.11 &   1.02 &   0.66 &   0.62 & \textbf{0.76} &   2.02 &  18.27 &  34.96 \\ 
					NN-ARDI &   0.99 &   0.62 &   1.36 &   1.10 &   1.68 &   2.44 &   8.11 &  69.17 & 136.08 &   0.78 &   1.05 &   1.26 &   0.64* &   0.60 &   0.81 &   1.62*** &   0.97 &   1.54 \\ 
					NN-ARDI+MARX &   1.14 &   1.27 &   1.03 &   1.24 &   2.78 &   9.83 &  61.72 & 143.26 & 261.02 &   2.55 &   1.71 &   2.03 &   0.71 &   1.04 &   0.87 &   2.76** &   2.66 &  14.35 \\ 
					\toprule					
				\end{tabular}
				\begin{tablenotes}[para,flushleft]
					Notes: See Table \ref{tab:all}.
				\end{tablenotes}
			\end{scriptsize}
		\end{threeparttable}
	\end{table}
\end{landscape}

\begin{landscape}
	\begin{table}[htp]
		\begin{threeparttable}
			\begin{scriptsize}
				\centering
				\footnotesize
				\caption{Quiet(er) Period (2011-2019)}
				\setlength{\tabcolsep}{0.5em} %
				\setstretch{1.45}
				\rowcolors{9}{gray!15}{white}
				\begin{tabular}{ l *{18}{l}}
					\toprule
					\hspace*{0.5cm} &
					\multicolumn{3}{c}{EMP} &
					\multicolumn{3}{c}{UNRATE} &
					\multicolumn{3}{c}{HOURS} & 
					\multicolumn{3}{c}{IP} &
					\multicolumn{3}{c}{IP MACH} &
					\multicolumn{3}{c}{RETAIL} \\
					
					\cmidrule(lr){2-4} \cmidrule(lr){5-7} \cmidrule(lr){8-10} \cmidrule(lr){11-13} \cmidrule(lr){14-16} \cmidrule(lr){17-19}
					& h=1 & h=2 & h=3 & h=1 & h=2 & h=3 & h=1 & h=2 & h=3 & h=1 & h=2 & h=3 & h=1 & h=2 & h=3 & h=1 & h=2 & h=3 \\ 
					\midrule
					RW &  1.47*** &  1.58* &  1.82 &  1.14*** &  1.28 &  1.35 &  1.11 &  1.13 &  1.20 &  1.04* &  1.05 &  1.04 &  1.04** &  1.07 & \textbf{0.91} &  1.22* &  1.30 &  1.33 \\ 
					ARDI,BIC &  1.19*** &  1.04 &  0.94 &  1.10* &  1.07 &  1.05 &  1.15 &  1.13 &  1.23 &  1.09 &  1.14 &  1.20 &  1.14 &  1.36 &  1.09 &  1.31 &  1.30 &  1.13 \\ 
					\midrule
					LASSO &  0.98 &  0.93 &  0.88 &  0.94 &  0.98 &  0.89 &  1.03 &  1.02 &  0.96 &  1.06 &  1.11 &  1.10 &  1.03 &  1.07 &  0.95 &  1.05* &  1.03 &  1.02 \\ 
					LASSO+MARX &  0.97 &  0.90 &  0.85 &  0.92 &  0.88 &  0.84 &  1.06 &  0.99 &  0.93 &  1.07 &  1.06 &  1.05 &  1.08 &  1.05 &  0.93 &  1.03 &  0.98 &  0.99 \\ 
					RIDGE &  0.98 & \textbf{ 0.87} &  0.83 &  0.85 &  0.83 &  0.80 &  1.05 &  0.99 &  0.95 &  1.05** &  1.06 &  1.05 &  1.03 &  1.06 &  0.92 &  1.12* &  1.07 &  0.96* \\ 
					RIDGE+MARX &  0.98 &  0.87 & \textbf{0.81} & \textbf{0.84} & \textbf{0.81} & \textbf{0.77} &  1.04 &  1.00 &  0.95 &  1.08 &  1.07 &  1.01 &  1.04 &  1.09 &  0.94 &  1.08 &  0.98 &  0.97 \\ 
					E-NET &  1.00 &  0.92 &  0.86 &  0.98 &  0.88 &  0.81 &  1.03 &  1.02 &  0.94 &  1.02 &  1.10 &  1.10 &  1.04 &  1.06 &  0.95 &  1.07 &  1.00 &  1.00 \\ 
					E-NET+MARX &  0.99 &  0.89 &  0.86 &  0.92 &  0.84 &  0.79 &  1.04 &  1.00 &  0.91 &  1.11 &  1.13 &  1.03 &  1.05 &  1.05 &  0.97 &  1.06 &  1.00 &  0.97** \\ 
					\midrule
					KRR-ARDI &  1.00** &  0.91 &  0.95 &  0.96* &  0.97 &  1.02 &  1.04 &  1.01 &  1.09 &  1.06* &  1.08 &  1.06 &  1.03** &  1.03 &  0.93 &  1.07* &  1.02 &  0.98 \\ 
					RF &  1.00 &  0.91 &  0.90 &  0.96** &  0.98* &  1.03 & \textbf{0.96} &  0.96 &  0.90 &  1.04 &  1.04 &  1.01 &  1.03 &  1.10 &  0.97 &  1.10 &  1.08 &  1.03 \\ 
					RF+MARX &  0.98 &  0.91** &  0.86** &  0.93*** &  0.92** &  0.94* &  1.00 &  0.97 & \textbf{0.89} &  1.07 &  1.01 &  1.00 &  1.02 &  1.09 &  0.99 &  1.10 &  1.06 &  1.02 \\ 
					Boosting &  0.98 &  0.90 &  0.88 &  0.98 &  1.02 &  1.05 &  0.98 &  0.95 &  0.92 &  1.06* &  1.06 &  1.04 &  1.03** &  1.09 &  0.93 &  1.06* &  1.05 &  0.97 \\ 
					Boosting+MARX &  0.99 &  0.90 &  0.87*** &  0.97 &  0.99 &  1.01 &  0.99 &  0.95 &  0.90 &  1.07* &  1.05 &  1.03 &  1.02** &  1.09 &  0.94 &  1.02* &  1.04 &  0.98 \\ 
					ARRF,2Ylag &  0.97 &  0.90** &  0.87* &  0.97* &  0.92* &  0.95 &  1.04 &  0.95 &  0.93 & \textbf{1.00} & \textbf{0.96} & \textbf{0.94} &  0.96 &  1.05 &  0.98 &  0.96 & \textbf{0.94} &  0.91 \\ 
					FA-ARRF,2Fac &  0.98 &  0.92 &  0.84 &  0.96 &  0.96 &  0.92 &  1.03 &  0.96 &  0.91 &  1.03 &  1.03 &  1.00 & \textbf{0.94} &  1.08 &  0.95 & \textbf{0.93} &  0.96 & \textbf{0.85} \\ 
					ARRF,6Ylag &  1.03 &  0.97 &  0.95 &  0.96 &  1.00 &  1.06 &  1.03 &  0.96 &  0.93 &  1.03 &  0.98 &  0.94 &  0.99 &  1.04 &  0.97 &  0.99 &  1.01 &  0.95 \\ 
					FA-ARRF,4Fac & \textbf{0.96} &  0.89 &  0.82 &  0.98 &  0.96 &  0.94 &  1.06 &  0.99 &  0.94 &  1.02 &  1.05 &  1.05 &  0.95 &  1.09 &  0.99 &  0.95 &  0.97 &  0.87 \\ 
					NN-ARDI &  1.00 &  0.99 &  0.98* &  1.06 &  1.10 &  1.04 &  1.02 & \textbf{0.93} &  0.95 &  1.05** &  1.01 &  1.03 &  0.98 & \textbf{1.01} &  0.94 &  1.02 &  1.05 &  0.99 \\ 
					NN-ARDI+MARX &  1.18** &  1.05 &  0.84 &  1.04 &  0.97 &  0.88 &  1.38 &  1.31 &  1.01 &  1.24 &  1.19 &  1.13 &  1.19 &  1.30 &  1.15 &  1.24 &  1.06 &  1.22 \\ 
					\toprule
					
				\end{tabular}
				\begin{tablenotes}[para,flushleft]
					Notes: See Table \ref{tab:all}.
				\end{tablenotes}
			\end{scriptsize}
		\end{threeparttable}
	\end{table}
\end{landscape}

\begin{landscape}
	\begin{scriptsize}
		\begin{table}[htp]
			\begin{threeparttable}
				\centering
				\footnotesize
				\caption{Quiet(er) Period (2011-2019), Continued}\label{tab:quiet}
				\setlength{\tabcolsep}{0.3202em} %
				\setstretch{1.45}
				\rowcolors{9}{gray!15}{white}
				\begin{tabular}{ l *{18}{l}}
					\toprule
					\hspace*{0.5cm} &
					\multicolumn{3}{c}{CPI} &
					\multicolumn{3}{c}{RPI} &
					\multicolumn{3}{c}{RPI HOUSING} &
					\multicolumn{3}{c}{CREDIT} &
					\multicolumn{3}{c}{HOUSE APP} &
					\multicolumn{3}{c}{PPI MANU} \\
					
					\cmidrule(lr){2-4} \cmidrule(lr){5-7} \cmidrule(lr){8-10} \cmidrule(lr){11-13} \cmidrule(lr){14-16} \cmidrule(lr){17-19}
					& h=1 & h=2 & h=3 & h=1 & h=2 & h=3 & h=1 & h=2 & h=3 & h=1 & h=2 & h=3 & h=1 & h=2 & h=3 & h=1 & h=2 & h=3 \\ 
					\midrule
					RW &  8.99*** & 11.14*** & 10.62*** &  2.34*** &  3.52*** &  4.29*** &  1.83* &  3.42 &  3.98 &  2.23 &  3.27 &  3.71 & \textbf{0.87} & \textbf{0.83} & \textbf{0.82} &  1.45*** &  1.26** &  1.24 \\ 
					ARDI,BIC &  1.66 &  1.83* &  1.79*** &  1.22*** &  1.42** &  1.50** &  2.42** &  5.50* &  6.65* &  1.52 &  1.53 &  1.58 &  1.17 &  1.36 &  1.41 &  1.49*** &  1.29** &  1.29** \\ 
					\midrule
					LASSO &  1.28 &  1.19 &  1.28 &  0.79 &  0.79 &  0.79 &  1.46 &  2.51 &  3.06 &  1.74 &  1.56 &  1.62 &  0.92 &  1.01 &  1.09 &  0.90 &  1.01 &  1.08 \\ 
					LASSO+MARX &  1.34 &  1.16 &  1.09* &  0.76* &  0.79 &  0.80 &  1.47 &  2.48 &  2.61 &  1.93 &  1.56 &  1.81 &  0.92 &  1.02 &  1.15 &  0.93 &  0.98 &  1.10 \\ 
					RIDGE &  1.21 &  1.29 &  1.32** &  0.78 &  0.81 &  0.86 &  1.53 &  2.89 &  3.33 &  1.50 &  1.48 &  1.62 &  1.03 &  1.05 &  1.13 &  1.13*** &  1.06** &  1.08** \\ 
					RIDGE+MARX &  1.09 &  1.13 &  1.14 & \textbf{0.75} &  0.74 &  0.81 &  1.35 &  2.99 &  3.52 &  1.61 &  1.59 &  1.58 &  1.08 &  1.14 &  1.19 &  1.04 &  1.04 &  1.07* \\ 
					E-NET &  1.26 &  1.13 &  1.11 &  0.80 &  0.78 &  0.81 &  1.36 &  2.37 &  3.11 &  1.79 &  1.53 &  1.69 &  0.91 &  1.01 &  1.13 &  0.92 &  0.98 &  1.10 \\ 
					E-NET+MARX &  1.24 &  1.19 &  1.13 &  0.78 &  0.78 &  0.80 &  1.32 &  2.60 &  2.87 &  1.66 &  1.49 &  1.69 &  0.91 &  1.04 &  1.10** &  0.95 &  1.00 &  1.10 \\ 
					\midrule
					KRR-ARDI &  1.00 &  0.94 &  0.93 &  0.84 &  0.87 &  0.90 &  1.51** &  3.38 &  4.28 &  1.30 &  1.29 &  1.25 &  1.13 &  1.29 &  1.26 &  0.94** & \textbf{0.90} & \textbf{0.92} \\ 
					RF &  1.15 &  1.25 &  1.27 &  0.80** &  0.78 &  0.82 &  0.72 &  0.57 &  0.65* &  0.94** &  0.91* &  1.02 &  0.92* &  1.03 &  1.06* &  0.98* &  1.00 &  1.03 \\ 
					RF+MARX &  1.08 &  1.19 &  1.22 &  0.80** &  0.78 &  0.83 &  0.81 &  0.63 &  0.78* &  0.97** &  0.93 &  0.98 &  0.94 &  1.15 &  1.31 &  1.06 &  1.06 &  1.11 \\ 
					Boosting &  1.51 &  1.66 &  1.62 &  0.83 &  0.82 &  0.82 &  0.83 &  0.72 &  0.81 &  1.02* &  1.06 &  1.16 &  0.90 &  0.94 &  1.00 &  1.04** &  1.05 &  1.03 \\ 
					Boosting+MARX &  1.46 &  1.65 &  1.62 &  0.82 &  0.79 &  0.81 &  0.81 &  0.70 &  0.77 &  1.02* &  1.05 &  1.14 &  0.92 &  0.98 &  1.00 &  1.03** &  1.04 &  1.03 \\ 
					ARRF,2Ylag &  0.93 &  0.98 &  0.98 &  0.87 &  0.84 &  0.87 &  0.38 &  0.52 &  0.61 &  1.07 &  1.00 &  1.07 &  0.97 &  0.95 &  0.96 &  0.94 &  0.98 &  1.05 \\ 
					FA-ARRF,2Fac & \textbf{0.91} &  1.04* &  1.00 &  0.79 &  0.72 &  0.74 &  0.57 &  1.03 &  1.22 &  1.20 &  1.21 &  1.26 &  1.06 &  1.12 &  1.17 &  0.91 &  0.97 &  1.03 \\ 
					ARRF,6Ylag &  0.91 & \textbf{0.91} & \textbf{0.90} &  0.89 &  0.88 &  0.86 & \textbf{0.29} & \textbf{0.46} & \textbf{0.46} & \textbf{0.93} & \textbf{0.81} & \textbf{0.74}* &  0.99 &  0.97 &  0.95 &  0.93 &  0.94 &  0.96 \\ 
					FA-ARRF,4Fac &  0.91 &  0.92* &  0.90 &  0.76 & \textbf{0.70} & \textbf{0.72} &  0.66 &  1.45 &  1.54 &  1.24 &  1.22 &  1.20 &  1.12 &  1.14 &  1.13 & \textbf{0.88} &  0.91 &  0.97 \\ 
					NN-ARDI &  1.40 &  1.22 &  1.18 &  0.85 &  0.89 &  0.95 &  0.95 &  1.40 &  1.91 &  1.44 &  1.47 &  1.33 &  0.96* &  1.05 &  0.94 &  1.17*** &  1.05 &  1.04 \\ 
					NN-ARDI+MARX &  1.15 &  1.51 &  1.18 &  1.00 &  0.86 &  0.86 &  1.66 &  2.80 &  2.97 &  2.02 &  2.33 &  2.02 &  1.30 &  1.28 &  1.31 &  1.14** &  1.09 &  1.21 \\ 
					\toprule				
				\end{tabular}
				\begin{tablenotes}[para,flushleft]
					Notes: See Table \ref{tab:all}.
				\end{tablenotes}
			\end{threeparttable}
		\end{table}
	\end{scriptsize}
\end{landscape}

\begin{landscape}
	\begin{table}[htp]
		\begin{threeparttable}
			\begin{scriptsize}
				\centering
				\footnotesize
				\caption{Pre-Covid (2008-2019)}
				\setlength{\tabcolsep}{0.5em} %
				\setstretch{1.45}
				\rowcolors{9}{gray!15}{white}
				\begin{tabular}{ l *{18}{l}}
					\toprule
					\hspace*{0.45cm} &
					\multicolumn{3}{c}{EMP} &
					\multicolumn{3}{c}{UNRATE} &
					\multicolumn{3}{c}{HOURS} & 
					\multicolumn{3}{c}{IP} &
					\multicolumn{3}{c}{IP MACH} &
					\multicolumn{3}{c}{RETAIL} \\
					
					\cmidrule(lr){2-4} \cmidrule(lr){5-7} \cmidrule(lr){8-10} \cmidrule(lr){11-13} \cmidrule(lr){14-16} \cmidrule(lr){17-19}
					& h=1 & h=2 & h=3 & h=1 & h=2 & h=3 & h=1 & h=2 & h=3 & h=1 & h=2 & h=3 & h=1 & h=2 & h=3 & h=1 & h=2 & h=3 \\ 
					\midrule
					RW & 1.33*** & 1.31* & 1.28 & 1.21*** & 1.21 & 1.17 & 1.00 & 1.00 & 1.01 & 0.98* & 0.96 & 0.92 & 0.96** & 1.00 & 0.93 & 1.15* & 1.17 & 1.07 \\ 
					ARDI,BIC & 1.32*** & 1.10 & 1.00 & 1.20* & 1.08 & 1.08 & 1.13 & 1.10 & 1.16 & 1.08 & 1.04 & 1.06 & 1.03 & 1.18 & 1.00 & 1.22 & 1.25 & 1.14 \\
					\midrule
					LASSO & 0.96 & 0.95 & 0.97 & 0.88 & 0.82 & 0.76 & 0.96 & 0.95 & 0.86 & 1.03 & 1.02 & 0.92 & 0.93 & 0.98 & 0.86 & 1.06* & 1.02 & 1.05 \\ 
					LASSO+MARX & 0.95 & 0.87 & 0.88 & 0.86 & 0.77 & 0.68 & 0.97 & 0.89 & 0.84 & 1.04 & 0.96 & 0.89 & 0.98 & 0.97 & 0.83 & 1.03 & 1.02 & 0.98 \\ 
					RIDGE & 0.95 & 0.84 & 0.77 & 0.81 & 0.68 & 0.61 & 0.96 & 0.91 & 0.85 & 0.95** & 0.91 & 0.88 & 0.92 & 0.92 & \textbf{0.81} & 1.12* & 1.10 & 0.96* \\ 
					RIDGE+MARX & \textbf{0.94} & \textbf{0.80} & \textbf{0.74} & \textbf{0.78} & \textbf{0.63} & \textbf{0.57} & 0.93 & 0.89 & \textbf{0.81} & 0.96 & \textbf{0.90} & 0.84 & 0.93 & \textbf{0.92} & 0.82 & 1.10 & 1.03 & 0.91 \\ 
					E-NET & 0.99 & 0.95 & 0.91 & 0.87 & 0.78 & 0.70 & 0.96 & 0.94 & 0.84 & 0.96 & 1.02 & 0.92 & 0.92 & 0.96 & 0.85 & 1.08 & 1.05 & 0.98 \\ 
					E-NET+MARX & 0.95 & 0.85 & 0.90 & 0.86 & 0.72 & 0.63 & 0.92 & 0.92 & 0.82 & 1.02 & 1.03 & 0.89 & 0.96 & 0.97 & 0.86 & 1.05 & 1.04 & 0.98** \\ 
					\midrule
					KRR-ARDI & 1.10** & 1.02 & 0.97 & 1.09* & 1.05 & 1.02 & 0.98 & 0.95 & 1.02 & 0.99* & 0.97 & 0.90 & 0.97** & 0.96 & 0.90 & 1.08* & 1.08 & 1.00 \\ 
					RF & 0.99 & 0.90 & 0.84 & 0.87** & 0.81* & 0.81 & 0.90 & 0.91 & 0.83 & 0.97 & 0.94 & 0.93 & 0.93 & 0.98 & 0.90 & 1.10 & 1.10 & 1.02 \\ 
					RF+MARX & 0.95 & 0.87** & 0.82** & 0.81*** & 0.74** & 0.79* & 0.93 & 0.91 & 0.83 & 0.97 & 0.93 & 0.94 & 0.93 & 0.97 & 0.91 & 1.07 & 1.09 & 1.02 \\ 
					Boosting & 1.03 & 0.95 & 0.90 & 0.97 & 0.94 & 0.94 & 0.92 & 0.90 & 0.86 & 0.98* & 0.97 & 0.92 & 0.94** & 0.99 & 0.90 & 1.09* & 1.10 & 1.00 \\ 
					Boosting+MARX & 1.02 & 0.93 & 0.87*** & 0.94 & 0.89 & 0.89 & 0.93 & 0.91 & 0.85 & 0.98* & 0.96 & 0.90 & 0.93** & 0.99 & 0.91 & 1.07* & 1.10 & 1.00 \\ 
					ARRF,2Ylag & 0.96 & 0.89** & 0.90* & 0.90* & 0.81* & 0.82 & 0.97 & 0.90 & 0.88 & 0.98 & 0.98 & 1.06 & 0.96 & 1.02 & 0.99 & 0.96 & \textbf{0.94} & 0.91 \\ 
					FA-ARRF,2Fac & 0.99 & 0.96 & 0.97 & 0.93 & 0.84 & 0.83 & 0.91 & \textbf{0.82} & 0.82 & \textbf{0.94} & 0.92 & 0.96 & \textbf{0.91} & 0.93 & 0.84 & \textbf{0.94} & 0.99 & 0.90 \\ 
					ARRF,6Ylag & 1.01 & 0.95 & 1.01 & 0.92 & 0.87 & 0.94 & 0.96 & 0.89 & 0.89 & 1.08 & 1.10 & 1.14 & 0.97 & 1.01 & 0.96 & 0.98 & 0.97 & 0.91 \\ 
					FA-ARRF,4Fac & 1.00 & 0.96 & 0.95 & 0.95 & 0.85 & 0.82 & 0.94 & 0.86 & 0.83 & 0.98 & 0.93 & 0.99 & 0.92 & 0.94 & 0.88 & 0.98 & 1.03 & \textbf{0.89} \\ 
					NN-ARDI & 1.05 & 0.96 & 0.91* & 1.06 & 0.89 & 0.83 & \textbf{0.90} & 0.88 & 0.87 & 0.95** & 0.97 & 0.92 & 0.93 & 0.93 & 0.91 & 1.04 & 1.07 & 0.98 \\ 
					NN-ARDI+MARX & 1.11** & 0.97 & 0.76 & 0.99 & 0.68 & 0.71 & 1.20 & 1.16 & 0.85 & 1.10 & 1.02 & \textbf{0.83} & 1.04 & 1.06 & 0.96 & 1.29 & 1.01 & 1.13 \\ 
					\toprule
				\end{tabular}
				\begin{tablenotes}[para,flushleft]
					Notes: See Table \ref{tab:all}.
				\end{tablenotes}
			\end{scriptsize}
		\end{threeparttable}
	\end{table}
\end{landscape}

\begin{landscape}
	\begin{scriptsize}
		\begin{table}[htp]
			\begin{threeparttable}
				\centering
				\footnotesize
				\caption{Pre-Covid (2008-2019), Continued}
				\setlength{\tabcolsep}{0.3464em} %
				\setstretch{1.45}
				\rowcolors{9}{gray!15}{white}
				\begin{tabular}{ l *{18}{l}}
					\toprule
					\hspace*{0.5cm} &
					\multicolumn{3}{c}{CPI} &
					\multicolumn{3}{c}{RPI} &
					\multicolumn{3}{c}{RPI HOUSING} &
					\multicolumn{3}{c}{CREDIT} &
					\multicolumn{3}{c}{HOUSE APP} &
					\multicolumn{3}{c}{PPI MANU} \\
					
					\cmidrule(lr){2-4} \cmidrule(lr){5-7} \cmidrule(lr){8-10} \cmidrule(lr){11-13} \cmidrule(lr){14-16} \cmidrule(lr){17-19}
					& h=1 & h=2 & h=3 & h=1 & h=2 & h=3 & h=1 & h=2 & h=3 & h=1 & h=2 & h=3 & h=1 & h=2 & h=3 & h=1 & h=2 & h=3 \\ 
					\midrule
					RW & 4.32*** & 5.34*** & 6.13*** & 1.88*** & 2.12*** & 2.37*** & 1.37* & 1.39 & 1.29 & 0.89 & 0.88 & 0.92 & 1.06 & 1.04 & 1.02 & 1.75*** & 1.45** & 1.30 \\ 
					ARDI,BIC & 1.78 & 1.82* & 1.85*** & 1.91*** & 1.93** & 1.66** & 2.66** & 2.86* & 2.91* & 1.02 & 1.00 & 1.02 & 1.35 & 1.70 & 1.67 & 2.58*** & 1.87** & 1.55** \\ 
					\midrule
					LASSO & 1.08 & 1.05 & 1.22 & \textbf{0.69} & 0.97 & 1.09 & 0.33 & \textbf{0.63} & 0.83 & 0.93 & 0.77 & \textbf{0.74} & 1.12 & 0.96 & 0.95 & 0.96 & 1.12 & 1.17 \\ 
					LASSO+MARX & 1.08 & 1.02 & 1.14* & 0.74* & 1.14 & 1.07 & \textbf{0.32} & 0.65 & \textbf{0.80} & 1.03 & 0.77 & 0.77 & 1.09 & 0.92 & 0.96 & 1.01 & 1.16 & 1.25 \\ 
					RIDGE & 1.00 & 1.08 & 1.30** & 0.88 & 1.05 & 1.20 & 0.79 & 0.93 & 0.94 & 0.86 & 0.76 & 0.76 & 0.99 & 0.93 & 0.91 & 1.23*** & 1.24** & 1.32** \\ 
					RIDGE+MARX & 0.97 & 1.01 & 1.15 & 0.77 & 1.02 & 1.11 & 0.60 & 0.88 & 0.86 & 0.94 & 0.86 & 0.87 & \textbf{0.99} & \textbf{0.86} & 0.86 & 1.11 & 1.12 & 1.27* \\ 
					E-NET & 1.02 & 1.03 & 1.16 & 0.83 & 0.95 & 1.09 & 0.35 & 0.64 & 0.85 & 0.95 & 0.76 & 0.77 & 1.02 & 0.93 & 0.92 & \textbf{0.95} & 1.17 & 1.27 \\ 
					E-NET+MARX & 1.08 & 1.08 & 1.09 & 0.79 & 0.98 & 1.08 & 0.36 & 0.68 & 0.82 & 0.99 & 0.78 & 0.78 & 1.05 & 0.91 & 0.92** & 1.01 & 1.10 & 1.27 \\ 
					\midrule
					KRR-ARDI & 0.91 & \textbf{0.87} & 0.97 & 1.03 & 1.03 & 1.08 & 1.50** & 1.66 & 1.62 & 0.91 & 0.90 & 0.93 & 1.11 & 1.19 & 1.21 & 1.29** & 1.20 & \textbf{1.15} \\ 
					RF & 0.90 & 0.95 & 1.14 & 0.92** & 1.02 & 1.16 & 0.82 & 1.05 & 1.15* & \textbf{0.80}** & \textbf{0.71}* & 0.77 & 1.08* & 1.18 & 1.41* & 1.20* & 1.35 & 1.45 \\ 
					RF+MARX & \textbf{0.88} & 0.90 & 1.08 & 0.90** & 1.04 & 1.21 & 0.72 & 0.99 & 1.12* & 0.82** & 0.74 & 0.78 & 1.07 & 1.21 & 1.44 & 1.14 & 1.29 & 1.43 \\ 
					Boosting & 1.07 & 1.11 & 1.23 & 1.03 & 1.04 & \textbf{1.07} & 1.15 & 1.28 & 1.28 & 0.80* & 0.77 & 0.81 & 1.05 & 1.05 & 1.09 & 1.29** & 1.25 & 1.19 \\ 
					Boosting+MARX & 1.05 & 1.09 & 1.20 & 1.01 & 1.03 & 1.08 & 1.13 & 1.27 & 1.27 & 0.81* & 0.77 & 0.82 & 1.05 & 1.05 & 1.08 & 1.27** & 1.24 & 1.20 \\ 
					ARRF,2Ylag & 1.85 & 1.14 & \textbf{0.97} & 1.02 & 1.04 & 1.20 & 1.20 & 0.96 & 1.08 & 0.97 & 0.97 & 1.02 & 1.02 & 0.98 & 1.13 & 1.00 & \textbf{1.08} & 1.36 \\ 
					FA-ARRF,2Fac & 1.54 & 1.19* & 1.30 & 0.77 & \textbf{0.92} & 1.09 & 0.84 & 0.94 & 0.84 & 0.98 & 0.96 & 1.03 & 1.14 & 0.96 & 1.25 & 1.08 & 1.18 & 1.60 \\ 
					ARRF,6Ylag & 1.60 & 1.25 & 1.22 & 1.05 & 1.22 & 1.33 & 1.22 & 0.97 & 1.56 & 1.17 & 1.12 & 1.27* & 1.03 & 1.20 & 1.54 & 1.12 & 1.22 & 1.63 \\ 
					FA-ARRF,4Fac & 1.59 & 1.28* & 1.43 & 0.70 & 0.95 & 1.10 & 0.88 & 1.02 & 1.05 & 1.02 & 0.96 & 0.98 & 1.13 & 1.04 & 1.09 & 1.04 & 1.19 & 1.55 \\ 
					NN-ARDI & 1.05 & 1.01 & 1.14 & 1.04 & 1.02 & 1.12 & 0.88 & 1.07 & 1.04 & 0.90 & 0.90 & 0.84 & 1.02* & 1.04 & 1.06 & 1.34*** & 1.20 & 1.20 \\ 
					NN-ARDI+MARX & 1.11 & 1.04 & 1.10 & 0.85 & 1.09 & 1.18 & 0.67 & 0.68 & 0.94 & 1.07 & 1.10 & 1.08 & 1.20 & 0.88 & \textbf{0.78} & 1.29** & 1.13 & 1.40 \\ 
					\toprule
				\end{tabular}
				\begin{tablenotes}[para,flushleft]
					Notes: See Table \ref{tab:all}.
				\end{tablenotes}
			\end{threeparttable}
		\end{table}
	\end{scriptsize}
\end{landscape}

\section{Additional Graphs}

\label{sec:addigraphs}

\begin{figure}[h!]
\caption{Variable Importance Measures for FA-ARRF(2,2) -- HOURS at $h=1$}
\label{fig:VI_hours}\centering
\includegraphics[width=0.85%
\textwidth]{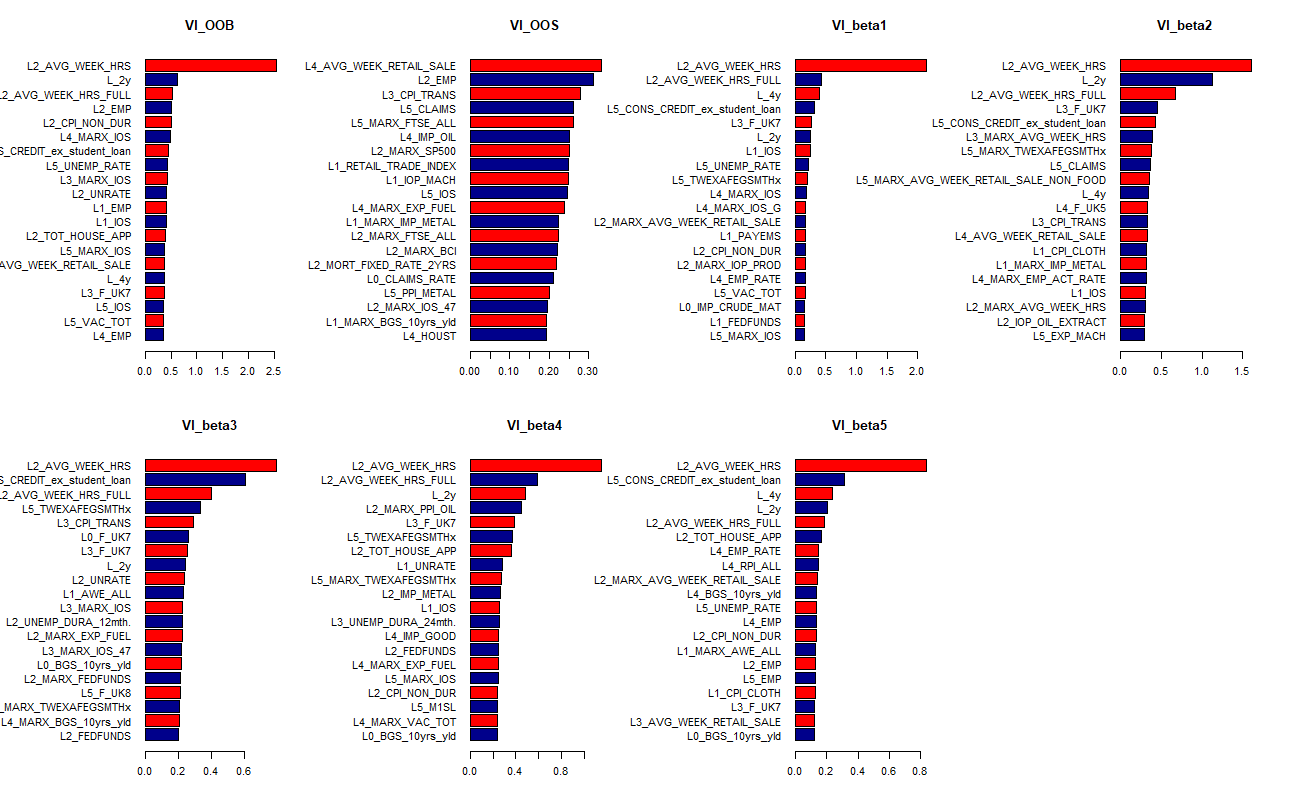} \flushleft
{\scriptsize \ \vspace{-2em} {\singlespacing
Notes: 20 most important series according to the various variable importance
(VI) criteria. Units are relative RMSE gains (in percentage) from including
the specific predictor in the forest part. $VI_{OOB}$ means VI for the
out-of-bag criterion. $VI_{OOS}$ is using the hold-out sample. $VI_{\beta}$
is an out-of-bag measure of how much $\beta_{t,k}$ varies by withdrawing a
certain predictor. }}
\par
{\scriptsize \ }
\end{figure}

\begin{figure}[h!]
\caption{Variable Importance Measures for ARRF(6) -- RPI HOUSE at $h=1$}
\label{fig:VI_RPIhouse}\centering
\includegraphics[width=0.85%
\textwidth]{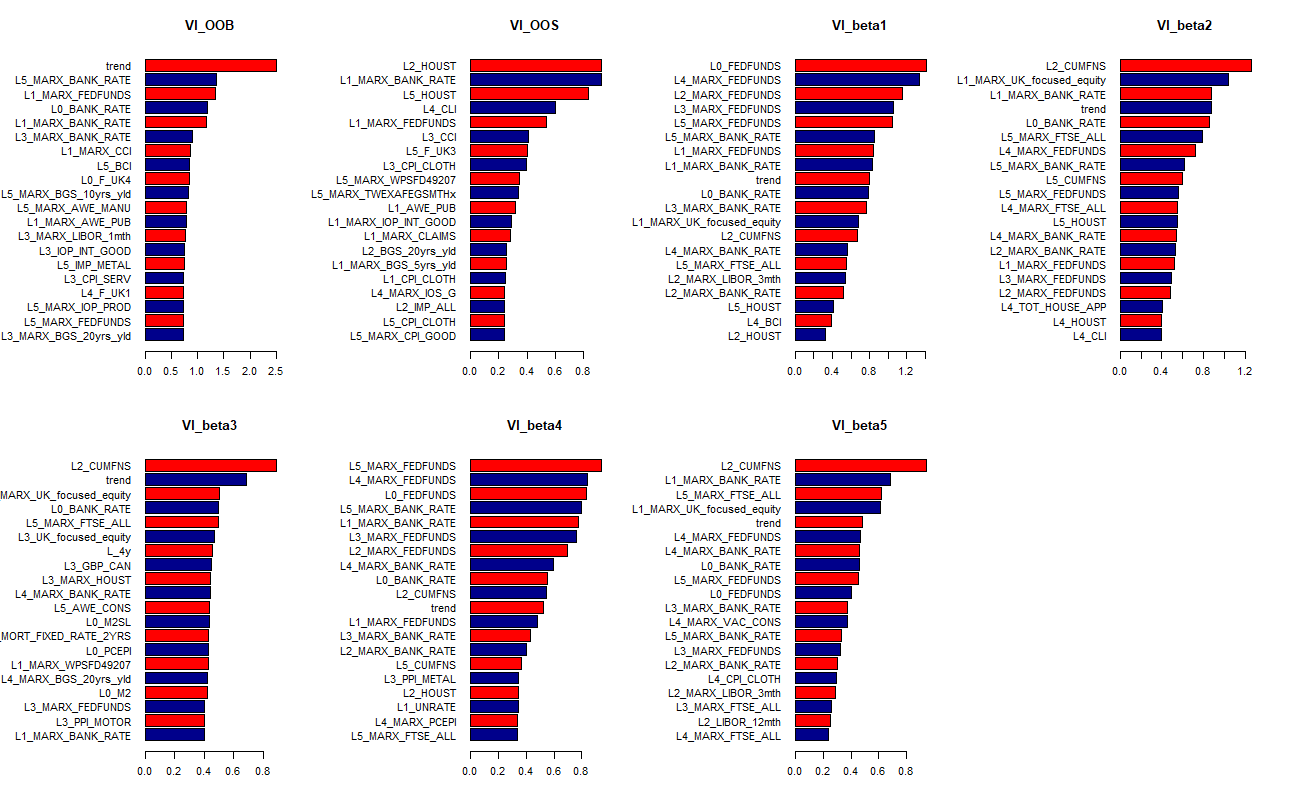} \flushleft
{\scriptsize \ \vspace{-2em} {\singlespacing
Notes: 20 most important series according to the various variable importance
(VI) criteria. Units are relative RMSE gains (in percentage) from including
the specific predictor in the forest part. $VI_{OOB}$ means VI for the
out-of-bag criterion. $VI_{OOS}$ is using the hold-out sample. $VI_{\beta}$
is an out-of-bag measure of how much $\beta_{t,k}$ varies by withdrawing a
certain predictor. }}
\par
{\scriptsize \ }
\end{figure}

\begin{figure}[h!]
\caption{Full POOS forecasts for RPI HOUSING at $h=1$}
\label{fig:forecasts_rpi}\centering
\includegraphics[width=0.65%
\textwidth]{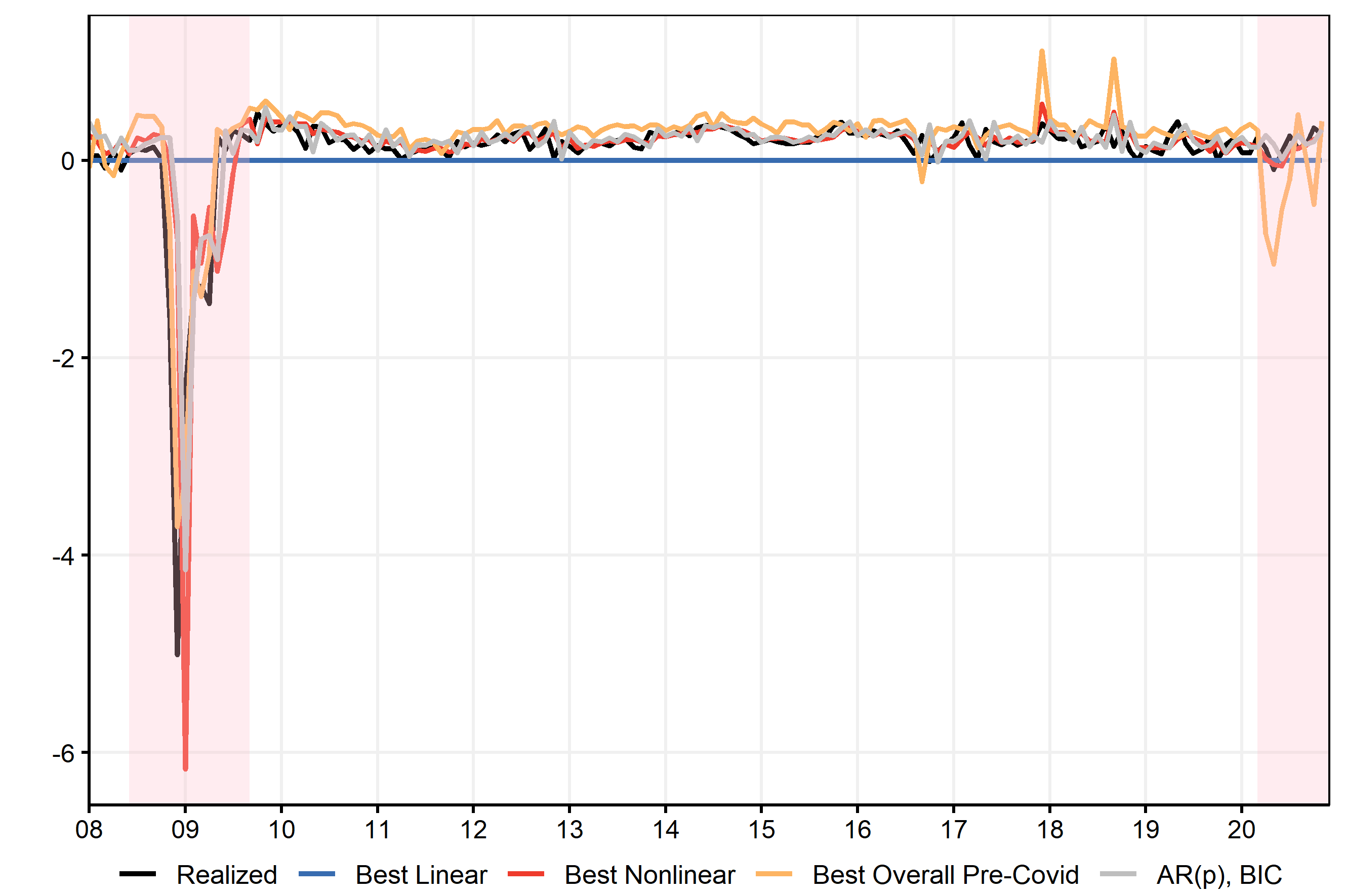} \flushleft
{\scriptsize \ \vspace{-2em} {\singlespacing
Notes: Pink shading corresponds to recessions. Exact selected models are
reported in Table \ref{quad}. }}
\par
{\scriptsize \ }
\end{figure}

\clearpage\onehalfspace


\section{UK Large Macroeconomic Dataset}\label{sec:UKdataappendix}

\noindent  When available, the series have been retrieved adjusted for seasonality beforehand. However, the price indices (CPI, RPI and PPI) were not and after conducting the \cite{kruskal1952use} test for seasonal behavior, these have been seasonally adjusted using the X-13-ARIMA-SEATS software developed by the United States Census Bureau. The transformation codes are: 1 - no transformation; 2 - first difference; 4 - logarithm; 5 - first difference of logarithm.  
\vspace{-.2cm}
\begin{tiny}
\begin{longtable}{lllllll}

 {\bf Id.} & {\bf Start date} & {\bf End date} & {\bf Variable} & {\bf Description}  & {\bf Source} & {\bf Code}  \\

\multicolumn{ 7}{c}{{\bf GROUP 1: LABOUR MARKET}} \\
1 & 71-02-01 & 20-09-01 & EMP & Number of People in Employment (aged 16 and over, seasonally adjusted) & ONS & 5 \\
2 & 92-04-01 & 20-09-01 & EMP\_PART & LFS: In employment: Part-time: UK: All: Thousands: SA & ONS & 5 \\
3 & 92-04-01 & 20-09-01 & EMP\_TEMP & LFS: Temporary employees: UK: All: Thousands: SA & ONS & 5 \\
4 & 71-02-01 & 20-09-01 & UNEMP\_RATE & Unemployment rate (aged 16 and over, seasonally adjusted) & ONS & 2 \\
5 & 92-04-01 & 20-09-01 & UNEMP\_DURA\_6mth & LFS: Unemployed up to 6 months: UK: All: Aged 16 and over: Thousands: SA & ONS & 5 \\
6 & 92-04-01 & 20-09-01 & UNEMP\_DURA\_6-12mth & LFS: Unemployed over 6 and up to 12 months: UK: All: Aged 16+: Thousands: SA & ONS & 5 \\
7 & 92-04-01 & 20-09-01 & UNEMP\_DURA\_12mth+ & LFS: Unemployed over 12 months: UK: All: Aged 16 and over: Thousands: SA & ONS & 5 \\
8 & 92-04-01 & 20-09-01 & UNEMP\_DURA\_24mth+ & LFS: Unemployed over 24 months: UK: All: Aged 16 and over: Thousands: SA & ONS & 5 \\
9 & 71-02-01 & 20-09-01 & EMP\_RATE & Employment rate (aged 16 to 64, seasonally adjusted) & ONS & 2 \\
10 & 71-02-01 & 20-09-01 & EMP\_ACT & LFS: Economically Active: UK: All: Aged 16-64: Thousands: SA & ONS & 5 \\
11 & 71-02-01 & 20-09-01 & EMP\_ACT\_RATE & LFS: Economic activity rate: UK: All: Aged 16-64 (\%): SA & ONS & 2 \\
12 & 71-01-01 & 20-11-01 & CLAIMS & Claimant Count : K02000001 UK : People : SA : Thousands & ONS & 5 \\
13 & 71-01-01 & 20-11-01 & CLAIMS\_RATE & Claimant Count : K02000001 UK : People : SA : Percentage (\%) & ONS & 2 \\
14 & 71-02-01 & 20-09-01 & TOT\_WEEK\_HRS & LFS: Total actual weekly hours worked (millions): UK: All: SA & ONS & 5 \\
15 & 92-04-01 & 20-09-01 & AVG\_WEEK\_HRS & LFS: Avg actual weekly hours of work: UK: All workers in main \& 2nd job: SA & ONS & 5 \\
16 & 92-04-01 & 20-09-01 & AVG\_WEEK\_HRS\_FULL & Average actual weekly hours of work for full-time workers (seasonally adjusted) & ONS & 5 \\
17 & 00-01-01 & 20-10-01 & AWE\_ALL & (Average Weekly Earning) AWE: Whole Economy Level : SA Total Pay Excluding Arrears & ONS & 5 \\
18 & 00-01-01 & 20-10-01 & AWE\_CONS & AWE: Construction Level : SA Total Pay Excluding Arrears & ONS & 5 \\
19 & 00-01-01 & 20-10-01 & AWE\_MANU & AWE: Manufacturing Level : SA Regular Pay Excluding Arrears & ONS & 5 \\
20 & 00-01-01 & 20-10-01 & AWE\_PRIV & AWE: Private Sector Level : SA Regular Pay Excluding Arrears & ONS & 5 \\
21 & 00-01-01 & 20-10-01 & AWE\_PUB & AWE: Public Sector Level : SA Total Pay Excluding Arrears & ONS & 5 \\
22 & 00-01-01 & 20-10-01 & AWE\_SERV & AWE: Services Level : SA Total Pay Excluding Arrears & ONS & 5 \\
23 & 75-02-01 & 20-10-01 & VAC\_TOT & UK Vacancies (thousands) - Total & FRED/ONS & 5 \\
24 & 01-05-01 & 20-10-01 & VAC\_CONS & UK Job Vacancies (thousands) - Construction & ONS & 5 \\
25 & 01-05-01 & 20-10-01 & VAC\_MANU & UK Job Vacancies (thousands) - Manufacturing & ONS & 5 \\

\multicolumn{ 7}{c}{{\bf GROUP 2: PRODUCTION}} \\
26 & 68-01-01 & 20-11-01 & IOP\_PROD & (Index of Production) IOP: B-E: PRODUCTION: CVMSA & ONS & 5 \\
27 & 95-01-01 & 20-11-01 & IOP\_CAP\_GOOD & IOP: MIG-CAG:Main Industrial Groupings - Capital Goods: CVMSA & ONS & 5 \\
28 & 95-01-01 & 20-11-01 & IOP\_DUR & IOP: MIG-CD:Main Industrial Groupings - Consumer Durables: CVMSA & ONS & 5 \\
29 & 95-01-01 & 20-11-01 & IOP\_ENER & IOP: MIG-NRG:Main Industrial Groupings - Energy: CVMSA & ONS & 5 \\
30 & 95-01-01 & 20-11-01 & IOP\_GOOD & IOP: MIG-COG:Main Industrial Groupings - Consumer Goods: CVMSA & ONS & 5 \\
31 & 95-01-01 & 20-11-01 & IOP\_INT\_GOOD & IOP: MIG-IG:Main Industrial Groupings - Intermediate Goods: CVMSA & ONS & 5 \\
32 & 68-01-01 & 20-11-01 & IOP\_MACH & IOP: CK:Manufacture of machinery and equipment n.e.c.: CVMSA & ONS & 5 \\
33 & 68-01-01 & 20-11-01 & IOP\_MANU & IOP: C:MANUFACTURING: CVMSA & ONS & 5 \\
34 & 68-01-01 & 20-11-01 & IOP\_MINE & IOP: B:MINING AND QUARRYING: CVMSA & ONS & 5 \\
35 & 95-01-01 & 20-11-01 & IOP\_NON\_DUR & IOP: MIG-CND:Main Industrial Groupings - Consumer Non-Durables: CVMSA & ONS & 5 \\
36 & 68-01-01 & 20-11-01 & IOP\_PETRO & IOP: CD:Manufacture of coke and refined petroleum product: CVMSA & ONS & 5 \\
37 & 95-01-01 & 20-11-01 & IOP\_OIL\_EXTRACT & IOP: 06:Extraction Of Crude Petroleum And Natural Gas: CVMSA & ONS & 5 \\

\multicolumn{ 7}{c}{{\bf GROUP 3: RETAIL AND SERVICES}} \\
38 & 97-01-01 & 20-11-01 & IOS & (Index of Services) IoS: Services: Index-1dp & ONS & 5 \\
39 & 97-01-01 & 20-11-01 & IOS\_45 & IoS: 45: Wholesale And Retail Trade And Repair Of Motor Vehicles And Motorcycles: Index-1dp & ONS & 5 \\
40 & 97-01-01 & 20-11-01 & IOS\_46 & IoS: 46: Wholesale trade except of motor vehicles and motorcycles: Index-1dp & ONS & 5 \\
41 & 97-01-01 & 20-11-01 & IOS\_47 & IoS: 47: Retail trade except of motor vehicles and motorcycles: Index-1dp & ONS & 5 \\
42 & 97-01-01 & 20-11-01 & IOS\_G & IoS: G: Wholesales, Retail and Motor Trade: Index-1dp & ONS & 5 \\
43 & 97-01-01 & 20-11-01 & IOS\_EDUC & IoS: O-Q: PAD, Education and Health Index-1dp & ONS & 5 \\
44 & 97-01-01 & 20-11-01 & IOS\_PNDS & IoS: H-N and R-U: PNDS: Private Non-Distribution Services: Index-1dp & ONS & 5 \\
45 & 96-01-01 & 20-11-01 & RSI & (Retail sales index) RSI:Volume Seasonally Adjusted:All Retailers inc fuel:All Business Index & ONS & 5 \\
46 & 60-01-01 & 20-11-01 & CAR\_REGIS & Sales: Retail trade: Car registration: Passenger cars for the United Kingdom, Number, SA & FRED & 5 \\
47 & 60-01-01 & 20-10-01 & RETAIL\_TRADE\_INDEX & Total Retail Trade in the United Kingdom, Index 2015=100, Monthly, SA & FRED & 5 \\
48 & 96-01-01 & 20-11-01 & AVGW\_RET\_SALE & All retailing including automotive fuel, VALUE SA - Average Weekly Retail Sales  & ONS & 5 \\
49 & 94-01-01 & 20-11-01 & AVGW\_RET\_SALE\_NF & Total retailing Predominantly non-food stores, VALUE SA - Average Weekly Retail Sales  & ONS & 5 \\

\multicolumn{ 7}{c}{{\bf GROUP 4: CONSUMER AND RETAIL PRICE INDICES}} \\
50 & 88-01-01 & 20-11-01 & CPIH\_ALL & CPIH INDEX 00: ALL ITEMS 2015=100, consumer price inflation incl. owner occupiers' housing costs (OOH) & ONS & 5 \\
51 & 88-01-01 & 20-11-01 & CPI\_ALL & CPI INDEX 00: ALL ITEMS 2015=100 & ONS & 5 \\
52 & 88-01-01 & 20-11-01 & CPI\_EX\_ENER & CPI INDEX: Excluding energy (SP) 2015=100 & ONS & 5 \\
53 & 88-01-01 & 20-11-01 & CPI\_GOOD & CPI INDEX: Goods 2015=100 & ONS & 5 \\
54 & 88-01-01 & 20-11-01 & CPI\_DUR & CPI INDEX: Durables (G) 2015=100 & ONS & 5 \\
55 & 88-01-01 & 20-11-01 & CPI\_NON\_DUR & CPI INDEX: Non-durables (G) 2015=100 & ONS & 5 \\
56 & 88-01-01 & 20-11-01 & CPI\_SERV & CPI INDEX: Services 2015=100 & ONS & 5 \\
57 & 88-01-01 & 20-11-01 & CPI\_CLOTH & CPI INDEX: Clothing \& footwear goods (G) 2015=100 & ONS & 5 \\
58 & 88-01-01 & 20-11-01 & CPI\_TRANS & CPI INDEX 07 : TRANSPORT 2015=100 & ONS & 5 \\
59 & 87-01-01 & 20-11-01 & RPI\_ALL & RPI All Items Index: Jan 1987=100 & ONS & 5 \\
60 & 87-01-01 & 20-11-01 & RPI\_GOOD & RPI: All Goods (Jan 1987=100) & ONS & 5 \\
61 & 87-01-01 & 20-11-01 & RPI\_SERV & RPI: All Services (Jan 1987=100) & ONS & 5 \\
62 & 87-01-01 & 20-11-01 & RPI\_HOUSE & RPI: Housing (Jan 1987=100) & ONS & 5 \\

\multicolumn{ 7}{c}{{\bf GROUP 5: INTERNATIONAL TRADE}} \\
63 & 97-01-01 & 20-11-01 & EXP\_TOT & Total Trade (TT): WW: Exports: BOP: CVM: SA & ONS & 5 \\
64 & 97-01-01 & 20-11-01 & EXP\_GOOD & Trade in Goods (T): WW: Exports: BOP: CVM: SA & ONS & 5 \\
65 & 97-01-01 & 20-11-01 & IMP\_ALL & Total Trade (TT): WW: Imports: BOP: CVM: SA & ONS & 5 \\
66 & 97-01-01 & 20-11-01 & IMP\_GOOD & Trade in Goods (T): WW: Imports: BOP: CVM: SA & ONS & 5 \\
67 & 97-01-01 & 20-11-01 & EXP\_FUEL & Trade in Goods: Fuels (3): WW: Exports: BOP: CVM: SA & ONS & 5 \\
68 & 97-01-01 & 20-11-01 & IMP\_FUEL & Trade in Goods: Fuels (3): WW: Imports: BOP: CVM: SA & ONS & 5 \\
69 & 97-01-01 & 20-11-01 & EXP\_OIL & Trade in Goods: Crude oil (33O): WW: Exports: BOP: CVM: SA & ONS & 5 \\
70 & 97-01-01 & 20-11-01 & IMP\_OIL & Trade in Goods: Crude oil (33O): WW: Imports: BOP: CVM: SA & ONS & 5 \\
71 & 97-01-01 & 20-11-01 & EXP\_MACH & Trade in Goods: Machinery and Transport (7): WW: Exports: BOP: CVM: SA & ONS & 5 \\
72 & 97-01-01 & 20-11-01 & IMP\_MACH & Trade in Goods: Machinery and Transport (7): WW: Imports: BOP: CVM: SA & ONS & 5 \\
73 & 97-01-01 & 20-11-01 & EXP\_METAL & Trade in Goods: Metal ores \& scrap (28): WW: Exports: BOP: CVM: SA & ONS & 5 \\
74 & 97-01-01 & 20-11-01 & IMP\_METAL & Trade in Goods: Metal ores \& scrap (28): WW: Imports: BOP: CVM: SA & ONS & 5 \\
75 & 97-01-01 & 20-11-01 & EXP\_CRUDE\_MAT & Trade in Goods: Crude Materials (2): WW: Exports: BOP: CVM: SA & ONS & 5 \\
76 & 97-01-01 & 20-11-01 & IMP\_CRUDE\_MAT & Trade in Goods: Crude Materials (2): WW: Imports: BOP: CVM: SA & ONS & 5 \\
77 & 80-01-01 & 20-12-01 & GBP\_BROAD & Monthly average Broad Effective exchange rate index, Sterling (Jan 2005 = 100)                          XUMABK82 & BOE & 5 \\
78 & 75-01-01 & 20-12-01 & GBP\_CAN & Monthly average Spot exchange rate, Canadian Dollar into Sterling                          XUMACDS & BOE & 5 \\
79 & 99-01-01 & 20-12-01 & GBP\_EUR & Monthly average Spot exchange rate, Euro into Sterling                          XUMAERS & BOE & 5 \\
80 & 75-01-01 & 20-12-01 & GBP\_JAP & Monthly average Spot exchange rate,  Japanese Yen into Sterling                          XUMAJYS & BOE & 5 \\
81 & 75-01-01 & 20-12-01 & GBP\_US & Monthly average Spot exchange rate, US\$ into Sterling                          XUMAUSS & BOE & 5 \\
82 & 87-06-01 & 20-12-01 & OIL\_PRICE & Crude Oil Prices: Brent - Europe, Dollars per Barrel, Monthly, NSA & BOE & 5 \\

\multicolumn{ 7}{c}{{\bf GROUP 6: MONEY, CREDIT AND INTEREST RATES}} \\
83 & 75-01-01 & 20-12-01 & BANK\_RATE & Monthly average of official Bank Rate              [a] [b]             IUMABEDR & BOE & 2 \\
84 & 93-04-01 & 20-11-01 & CONS\_CREDIT & Monthly amounts outstanding of total (excluding the Student Loans Company) sterling consumer credit & & \\
& & & & lending to individuals (in sterling millions) SA   & BOE & 5 \\
85 & 97-10-01 & 20-11-01 & TOT\_LENDING\_APP & Monthly number of total sterling approvals for secured lending to individuals SA          & BOE & 5 \\
86 & 93-04-01 & 20-11-01 & TOT\_HOUSE\_APP & Monthly number of total sterling approvals for house purchase to individuals SA       & BOE & 5 \\
87 & 95-01-01 & 20-12-01 & MORT\_FRATE\_5YRS & Monthly interest rate of UK monetary financial institutions (excl. Central Bank) sterling 5 year (75\% LTV) & & \\ 
& & & & fixed rate mortgage to households (in percent) NSA      & BOE & 2 \\
88 & 95-01-01 & 20-12-01 & MORT\_FRATE\_2YRS & Monthly interest rate of UK monetary financial institutions (excl. Central Bank) sterling 2 year (75\% LTV) & & \\ 
& & & & fixed rate mortgage to households (in percent) NSA       & BOE & 2 \\
89 & 86-09-01 & 20-11-01 & M1 & Monthly amounts outstanding of monetary financial institutions' sterling and all foreign currency M1  & & \\ 
& & & & (UK estimate of EMU aggregate) liabilities to private and public sectors (in sterling millions) SA      & BOE & 5 \\
90 & 86-12-01 & 20-11-01 & M2 & Monthly amounts outstanding of monetary financial institutions' sterling and all foreign currency M2  & & \\ 
& & & & (UK estimate of EMU aggregate) liabilities to private and public sectors (in sterling millions) SA      & BOE & 5 \\
91 & 87-01-01 & 20-11-01 & M3 & Monthly amounts outstanding of monetary financial institutions' sterling and all foreign currency M3  & & \\ 
& & & & (UK estimate of EMU aggregate) liabilities to private and public sectors (in sterling millions) SA     & BOE & 5 \\
92 & 82-06-01 & 20-09-01 & M4 & Monthly amounts outstanding of M4 (monetary financial institutions' sterling M4 liabilities to private sector) & & \\ 
& & & & (in sterling millions) SA  & BOE & 5 \\
93 & 86-01-01 & 20-12-01 & LIBOR\_1mth & 1-Month London Interbank Offered Rate (LIBOR), based on British Pound, Percent, Monthly, NSA & FRED & 2 \\
94 & 86-01-01 & 20-12-01 & LIBOR\_3mth & 3-Month London Interbank Offered Rate (LIBOR), based on British Pound, Percent, Monthly, NSA & FRED & 2 \\
95 & 86-01-01 & 20-12-01 & LIBOR\_12mth & 12-Month London Interbank Offered Rate (LIBOR), based on British Pound, Percent, Monthly, NSA & FRED & 2 \\
96 & 93-12-01 & 20-12-01 & BGS\_5yrs\_yld & Monthly average yield from British Government Securities, 5 year Nominal Par Yield          & BOE & 2 \\
97 & 93-12-01 & 20-12-01 & BGS\_10yrs\_yld & Monthly average yield from British Government Securities, 10 year Nominal Par Yield      & BOE & 2 \\
98 & 00-01-01 & 20-12-01 & BGS\_20yrs\_yld & Monthly average yield from British Government Securities, 20 year Nominal Par Yield             & BOE & 2 \\

\multicolumn{ 7}{c}{{\bf GROUP 7: STOCK MARKET}} \\
99 & 80-02-01 & 20-12-01 & FTSE\_ALL & UK FTSE All Share (FTAS) & YAHOO & 5 \\
100 & 85-12-01 & 20-12-01 & FTSE250 & FTSE 250 (FTMC) & YAHOO & 5 \\
101 & 90-01-01 & 20-12-01 & VIX & CBOE Volatility Index (VIX) & YAHOO & 1 \\
102 & 60-01-01 & 20-12-01 & SP500 & S\&P 500 (GSPC) & YAHOO & 5 \\
103 & 96-03-01 & 20-12-01 & UK\_focused\_equity & iShares MSCI United Kingdom ETF (EWU) & YAHOO & 5 \\
104 & 87-01-01 & 20-12-01 & EUR\_UNC\_INDEX & Economic Policy Uncertainty Index for Europe, Index, Monthly, NSA & FRED & 2 \\

\multicolumn{ 7}{c}{{\bf GROUP 8: SENTIMENT AND LEADING INDICATORS}} \\
105 & 77-03-01 & 20-11-01 & BCI & Business confidence index (BCI)Amplitude adjusted, Long-term average = 100 & OECD & 2 \\
106 & 74-01-01 & 20-12-01 & CCI & Consumer confidence index (CCI)Amplitude adjusted, Long-term average = 100 & OECD & 2 \\
107 & 60-01-01 & 20-12-01 & CLI & Composite leading indicator (CLI)Amplitude adjusted, Long-term average = 100 & OECD & 2 \\

\multicolumn{ 7}{c}{{\bf GROUP 9: PRODUCER PRICE INDICES}} \\
108 & 60-01-01 & 20-11-01 & PPI\_MANU & Producer price indices (PPI)Manufacturing, domestic market, 2015=100 & OECD & 5 \\
109 & 96-01-01 & 20-11-01 & PPI\_MACH & PPI Machinery and Equipment N.E.C. for Domestic Market (G6VG) & ONS & 5 \\
110 & 96-01-01 & 20-11-01 & PPI\_OIL & PPI Coke and Refined Petroleum Products for Domestic Market (G6ST) & ONS & 5 \\
111 & 96-01-01 & 20-11-01 & PPI\_METAL & PPI Basic Metals for Domestic Market (G6SZ) & ONS & 5 \\
112 & 96-01-01 & 20-11-01 & PPI\_MOTOR & PPI Motor Vehicles, Trailers and Semi-Trailers for Domestic Market (G6WH) & ONS & 5 \\

\end{longtable}

\end{tiny}

\clearpage
\section{US Data}\label{sec:USdataappendix}

\noindent The additional transformation codes are: 6 - second difference of logs; 7 - $\delta (x_t / x_{t-1} - 1)$.  

\begin{tiny}
\begin{flushleft}
\begin{longtable}{lllllc}

{\bf Start date} & {\bf End date} & {\bf Variable} & {\bf Description}  & {\bf Source} & {\bf Code}  \\

98-01-01 & 20-11-01 & W875RX1 & Real personal income ex transfer receipts & FREDMD & 5 \\
98-01-01 & 20-11-01 & INDPRO & IP Index & FREDMD & 5 \\
98-01-01 & 20-11-01 & CUMFNS & Capacity Utilization: Manufacturing & FREDMD & 2 \\
98-01-01 & 20-11-01 & UNRATE & Civilian Unemployment Rate & FREDMD & 2 \\
98-01-01 & 20-11-01 & PAYEMS & All Employees: Total nonfarm & FREDMD & 5 \\
98-01-01 & 20-11-01 & CES0600000008 & Avg Hourly Earnings : Goods-Producing & FREDMD & 6 \\
98-01-01 & 20-11-01 & HOUST & Housing Starts: Total New Privately Owned & FREDMD & 4 \\
98-01-01 & 20-11-01 & DPCERA3M086SBEA & Real personal consumption expenditures & FREDMD & 5 \\
98-01-01 & 20-11-01 & CMRMTSPLx & Real Manu. and Trade Industries Sales & FREDMD & 5 \\
98-01-01 & 20-11-01 & M1SL & M1 Money Stock & FREDMD & 6 \\
98-01-01 & 20-11-01 & M2SL & M2 Money Stock & FREDMD & 6 \\
98-01-01 & 20-11-01 & TOTRESNS & Total Reserves of Depository Institutions & FREDMD & 6 \\
98-01-01 & 20-11-01 & NONBORRES & Reserves Of Depository Institutions & FREDMD & 7 \\
98-01-01 & 20-11-01 & FEDFUNDS & Effective Federal Funds Rate & FREDMD & 2 \\
98-01-01 & 20-11-01 & GS10 & 10-Year Treasury Rate & FREDMD & 2 \\
98-01-01 & 20-11-01 & TWEXAFEGSMTHx & Trade Weighted U.S. Dollar Index & FREDMD & 5 \\
98-01-01 & 20-11-01 & WPSFD49207 & PPI: Finished Goods & FREDMD & 6 \\
98-01-01 & 20-11-01 & CPIAUCSL & CPI : All Items & FREDMD & 6 \\
98-01-01 & 20-11-01 & PCEPI & Personal Cons. Expend.: Chain Index & FREDMD & 6 \\

\end{longtable}
\end{flushleft}

\end{tiny}

\end{document}